\documentclass[12pt,onecolumn,draftclsnofoot]{IEEEtran}
\ifCLASSOPTIONcompsoc
  \usepackage[nocompress]{cite}
\else
  \usepackage{cite}
\fi
\ifCLASSINFOpdf
\else
\fi
\usepackage{amsmath,amsfonts,amssymb}
\usepackage{algorithmic}
\usepackage{algorithm}
\usepackage{array}
\usepackage[caption=false,font=normalsize,labelfont=sf,textfont=sf]{subfig}
\usepackage{textcomp}
\usepackage{stfloats}
\usepackage{url}
\usepackage{verbatim}
\usepackage{graphicx}
\usepackage{cite}
\usepackage{xcolor}
\usepackage{tikz}

\newtheorem{theorem}{Theorem}
\newtheorem{lemma}{Lemma}
\newtheorem{remark}{Remark}
\newtheorem{assume}{Assumption}
\newtheorem{define}{Definition}

\usepackage[switch,columnwise]{lineno}

\begin{document}

\title{Randomized Matrix Weighted Consensus}
\author{Nhat-Minh Le-Phan,~Minh Hoang Trinh$^{\ast}$,~\IEEEmembership{Member,~IEEE},~Phuoc Doan Nguyen
\IEEEcompsocitemizethanks{\IEEEcompsocthanksitem N.-M Le-Phan and P. D. Nguyen are with Department of Automation Engineering, School of Electrical and Electronic Engineering, Hanoi University of Science and Technology (HUST), Hanoi, Vietnam. N.-M. Le-Phan is also with Navigation, Guidance and Control Technologies Center, Viettel Aerospace Institute, Hanoi, Vietnam. E-mails: minh.lpn221013m@sis.hust.edu.vn; phuoc.nguyendoan@hust.edu.vn. M. H. Trinh is with AI Department, FPT University, Binh Dinh, Vietnam. Email: minhtrinh@ieee.org. Corresponding author: M.~H.~Trinh.\protect}
\thanks{Manuscript received \ldots}
}

\IEEEtitleabstractindextext{%
\begin{abstract}
In this paper, randomized gossip-type matrix weighted consensus algorithms are proposed for both leaderless and leader-follower topologies. {First, we introduce the notion of expected matrix weighted network, which captures the multi-dimensional interactions between any two agents in a probabilistic sense.} Under some mild assumptions on the distribution of the expected matrix weights and the upper bound of the updating step size, the proposed asynchronous pairwise update algorithms {drive the network to achieve a consensus in expectation.} An upper bound of the $\epsilon$-convergence time of the algorithm is then derived. Furthermore, the proposed algorithms {are applied to} the bearing-based network localization and formation control problems. The theoretical results are supported by several numerical examples.
\end{abstract}

}

\maketitle

\IEEEdisplaynontitleabstractindextext

%
\IEEEpeerreviewmaketitle

\section{Introduction}
\label{sec:introduction}
\IEEEPARstart{I}{n} recent years, dynamics on complex networks have received a considered amount of research attention \cite{Newman2018networks}. Particularly, consensus algorithms have been shown to be essential for various applications \cite{Olfati2007consensuspieee} such as formation control, network localization, distributed estimation, synchronization, and social networks, \ldots

The dynamics of the agents in a network are often described by vectors of more than one state variable, and there may exist cross-layer couplings between these state variables. The authors in \cite{THM2018} proposed a matrix-weighted consensus algorithm, in which each pair-wise interaction between neighboring agents is associated with a symmetric nonnegative matrix weight. {Matrix-weighted network, thus, provides a model for studying dynamics processes on a diffusive multi-dimensional network.} Corresponding to a matrix weighted network, we can define a matrix weighted graph and study its algebraic structure. {Unlike scalar-weighted networks, topology connectedness of a matrix weighted network does not guarantee the agents to asymptotically converge to a common value under the consensus algorithm.} Algebraic algorithms to determine whether the system would achieve consensus or clustering behaviors were considered in \cite{THM2018,Kwon2020matrix,Barooah2006graph,Pan2020controllability,Nguyen2022MCA}. Matrix weighted consensus algorithms can be found in modeling multi-dimensional opinion dynamics \cite{Ahn2020opinion,Pan2018bipartite}, formation control and network localization \cite{bearingzhao,zhao2016aut,Barooah2006graph,Oh2012TAC}, and the synchronization of multi-dimensional coupled mechanical and electrical systems \cite{Tuna2016aut,Tuna2019synchronization,Li2020synchronization}. 

{The pioneering work on matrix weighted consensus is attributed to \cite{THM2018}, where the authors proposed consensus algorithms and conditions for achieving global or clustered consensus.}. Continuous-time matrix weighted consensus with switching graph topologies was studied in \cite{Pan2021consensus}. Discrete-time matrix weighted consensus with fixed- or switching topologies was studied in \cite{quoc2021}, while hybrid continuous-discrete time update algorithms were proposed in \cite{Miao2022TNSE,Miao2022consensus}. {The event-trigger consensus of matrix weighted network was studied in \cite{Pan2023TNSE}. In \cite{THM2022CDC}, the matrix-scaled consensus algorithm was proposed for both single and double-integrator agents. It is noteworthy that most existing works on matrix-weighted consensus assume that the agents update their states simultaneously and according to a deterministic update sequence, which causes expensive communication costs and computational burden.}

{Randomized consensus algorithms refer to a family of consensus algorithms in which a subgroup of agents are randomly selected for updating information at each discrete instant according to some stochastic model \cite{FB-LNS}.} These algorithms, with their immense simplicity and reliability, have been widely {implemented in different applications in wireless sensor networks.} The randomized gossip algorithm proposed in \cite{boyd2006} was a significant step toward solving the average consensus problem in a randomized manner. In this algorithm, a pair of agents $i$ and $j$ are said to ``gossip'' at a discrete time instant $k$ if they set their next values as the average of their current ones. {Each agent is associated with a clock} that ``ticks'' are distributed as a rate 1 Poisson process. In other words, at each time slot, there will be one agent $i$ who wakes up with a probability $\frac{1}{n}$, and it will choose another agent $j$ with a probability ${\rm P}_{ij}$ (conditional to $i$ being selected) to gossip. Many {extensions} have been proposed based on \cite{boyd2006} to enhance its convergence speed{, capability} as well as robustness, see for example, \cite{5625615,TIT2010,1238221} on geographic gossip; \cite{HE20118718,7887729,LABAHN1994235} on periodic gossiping; \cite{7581065,4787122} on broadcast gossip; \cite{8946047} on reducing the probability of selecting duplicate nodes,\dots However, these studies only cover the situation when the connecting weights between agents are scalar, i.e., they do not have the ability to deal with the coupling mechanism for agents with multi-dimensional state vectors.

{The contributions of this paper are summarized as follows. First, we define the notion of expected matrix-weighted network and expected matrix-weighted Laplacian. Based on the newly defined notions, we propose randomized matrix-weighted consensus algorithms for networks with a leaderless or a leader-follower structure. The proposed algorithms can be considered as adaptations of the pair-wise gossiping update protocol in \cite{boyd2006} for matrix-weighted networks. Second, the convergence process of the proposed algorithms are examined. To guarantee consensus in expectation in a scalar-weighted network, most existing works \cite{boyd2006,5340530} specified a condition related to the connectivity of the network. In contrast, the condition for achieving consensus in expectation in a matrix-weighted graph is jointly determined by both the eigenvalues of the expected graph Laplacian and the singular values of the local matrix weights of each agent. Lastly, we demonstrate two applications of the proposed algorithms in bearing-based network localization and formation control problems. Although there were a considered amount of existing works on matrix weighted consensus, to the best of our knowledge, this is the first work considering randomized asynchronous updates on a directed matrix-weighted network. It is also worth noting that randomization may symmetry a directed matrix weighted graph, so that the convergence analysis can be performed for a corresponding undirected matrix-weighted graph.}

The rest of this paper is organized as follows. In Section~\ref{sec:prel}, we introduce the preliminaries and problem formulation. Our works on randomized matrix weighted consensus algorithms for leaderless and leader-follower topologies are covered in Sections~\ref{sec:leaderless} and \ref{sec:leader-follower}, respectively. A randomized bearing-based network localization algorithm is presented in Section~\ref{section: NL}.  {black}{Section~\ref{section: FC} extends our work to the formation control problem based on position estimation.} Numerical simulations are provided in Section~\ref{section: simulation} to support the theoretical results. Finally, we summarize the paper and outline directions for future research in Section~\ref{sec:conclusion}.

\section{Preliminaries and problem formulation}
\label{sec:prel}
{
In this section, we begin by presenting the definition of matrix weighting, a concept referenced in \cite{THM2018}. Following this, we introduce the concept of the expected graph. This particular concept, which is one of the contributions of this article, will play a crucial role in our subsequent analyses.  Finally, we present the gossip-based communication protocol utilized in this paper.}
\subsection{Matrix weighted graphs and the expected graphs}
A directed matrix weighted graph is denoted by $ \mathcal{G}=(V,E,A)$, where, $V=\left\{ 1,2,...,n\right\}$ is the vertex set (agents), $E\subseteq V\times V$ is the edge set, and 
$A=\{ \mathbf{A}_{ij} \in \mathbb{R}^{d \times d}|~(i,j) \in E\}$ denotes the set of matrix weights. $d \geq 1$ is the dimension of each agent's state vector. When $d = 1$, $\mathcal{G}$ reduces to a scalar graph. The interactions between any two agents in $\mathcal{G}$ are captured by the corresponding matrix weights: if $i$ can communicate with $j$, there is a symmetric positive definite/positive semi-definite matrix weight $\mathbf{A}_{ij}=\mathbf{A}_{ij}^\top\geq 0$; and if $i$ does not have a communication link with $j$, then  $\mathbf{A}_{ij}$ is the zero matrix. We call an edge $(i,j)$ positive definite (resp., positive semi-definite) if its corresponding matrix weight is positive definite (resp., positive semi-definite). {The \emph{matrix weighted adjacency matrix} of $\mathcal{G}$ is defined as $\mathbf{A}=[\mathbf{A}_{ij}]_{i,j=1\dots n}$}. Let $\mathbf{D}_{i}= \sum_{j \in V} \mathbf{A}_{ij}$ be the degree of vertex $i$, and {$\mathbf{D} = \text{blkdiag}\left\{\mathbf{D}_{1}, \ldots,\mathbf{D}_{n}\right\}$} denote \emph{degree matrix} of $\mathcal{G}$. The \emph{matrix weighted Laplacian} is then defined as $\mathbf{L} = \mathbf{D} - \mathbf{A} \in \mathbb{R}^{nd \times nd}$. 

\begin{figure}
    \centering
    \includegraphics[width=\linewidth]{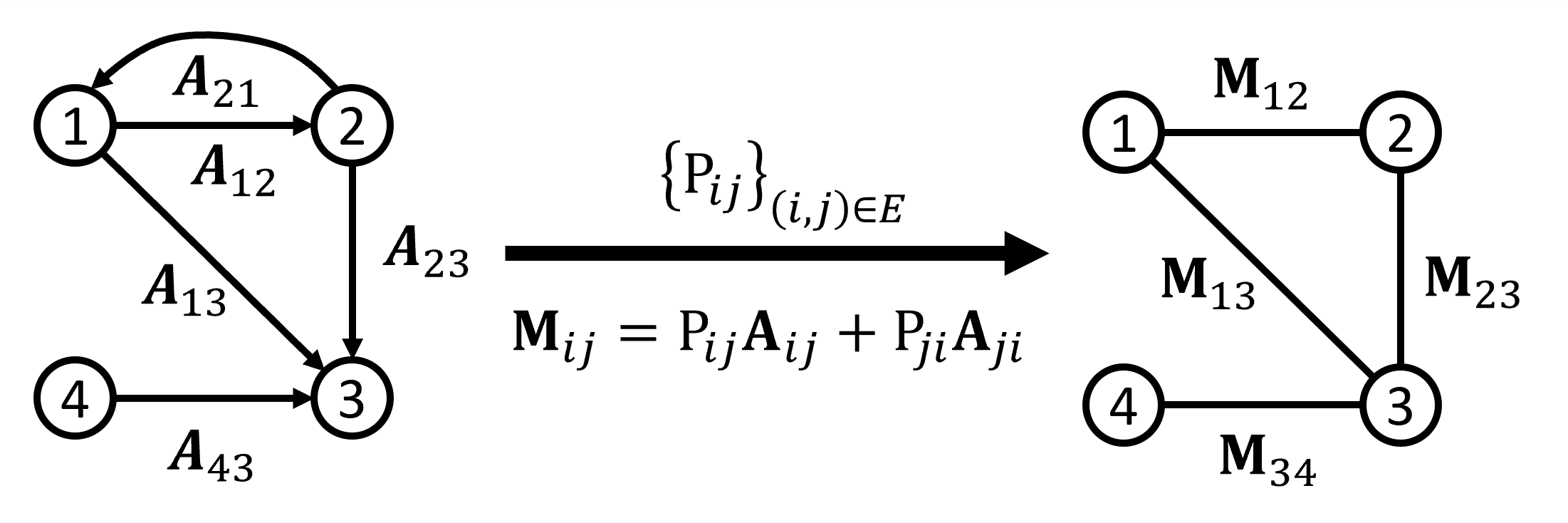}
    \caption{A directed matrix weighted graph $\mathcal{G}$ and its corresponding expected graph $\mathcal{G}^{\rm M}$.}
    \label{fig:graphSymmetrization}
\end{figure}
For each edge in $E$, we associate a corresponding number ${\rm P}_{ij}\in [0,1]$, which is the probability that a vertex $j \in V$ is chosen by vertex $i$ for updating information, $\sum_{j=1}^n{\rm P}_{ij}=1$. Defining $\mathbf{M}_{ij}=\frac{1}{n}(\mathbf{A}_{ij} {\rm P}_{ij}+\mathbf{A}_{ji} {\rm 
 P}_{ji})$ as the \emph{expected matrix weight} between two vertices $i$ and $j$, it is obvious that $\mathbf{M}_{ij}=\mathbf{M}_{ji}=\mathbf{M}_{ij}^{\top} \geq 0$. Thus, the digraph $\mathcal{G}$ induces a corresponding \emph{undirected} matrix weighted graph $\mathcal{G}^{\rm M}=(V,E^{\rm M},A^{\rm M})$, where $E^{\rm M}=\{(i,j)|~\mathbf{M}_{ij} \neq 0)\}$ and $A^{\rm M}=\{ \mathbf{M}_{ij}\}_{(i,j)\in E^{\rm M}}$. {We call $\mathcal{G}^{\rm M}$ the \emph{expected matrix-weighted graph}, or the expected graph  of $\mathcal{G}$. For example, Fig. ~\ref{fig:graphSymmetrization} depicts a matrix weighted graph $\mathcal{G}$ and its expected graph $\mathcal{G}^{\rm M}$. }
 
  {The \emph{expected adjacency matrix},  \emph{expected degree matrix}, and \emph{expected Laplacian matrix} of $\mathcal{G}$, denoted as $\mathbf{A}^{\rm M}$, $\mathbf{D}^{\rm M}$, and $\mathbf{L}^{\rm M} \in \mathbb{R}^{nd \times nd}$, respectively, are defined as the matrix weighted adjacency, degree, and Laplacian matrix of expected graph $\mathcal{G}^{\rm M}$. In other words, $\mathbf{A}^{\rm M}=[\mathbf{M}_{ij}]_{i,j=1,\dots,n}$, $\mathbf{D}^{\rm M}=\text{blkdiag}(\mathbf{D}^M_1,\dots,\mathbf{D}^M_n)$ where $\mathbf{D}^M_i=\sum_{j \in V}\mathbf{M}_{ij}$, and $\mathbf{L}^{\rm M} = \mathbf{D}^{\rm M}-\mathbf{A}^{\rm M} \in \mathbb{R}^{nd \times nd}$.} Potentially, all existing results on undirected matrix weighted graphs can be used for studying the expected graph $\mathcal{G}^{\rm}$. A positive path $\mathcal{P}=i_1,i_2,...,i_l \in \mathcal{G}^{\rm M}$ is a sequence of non-repeating edges in $V$ such that $\mathbf{M}_{i_{k}i_{k+1}} > 0, k=1,...,l-1$. 
A positive spanning tree $\mathcal{T}^{\rm M}$ is a graph consisting of $n$ vertices and $(n-1)$ edges in $\mathcal{G}^{\rm M}$ so that there exists a unique positive path between any pair of vertices in $\mathcal{T}^{\rm M}$.
 
\begin{lemma}\label{nullspace}
\emph{\cite{THM2018}}
The expected Laplacian matrix $\mathbf{L}^{\rm M}$ is symmetric and positive semi-definite. Moreover, 
  ${\rm null}(\mathbf{L}^{\rm M}) = {\rm span}\{ {\rm range}(\mathbf{1}_n \otimes \mathbf{I}_d), \{ \mathbf{v}=[v_1^\top,\dots, v_n^\top]^\top \in \mathbb{R}^{nd}|~(v_j-v_j)\in {\rm null}(\mathbf{M}_{ij}), \forall (i,j) \in E \} \}$.
\end{lemma}
\subsection{Randomized distributed protocol}
In this subsection, we describe the basic protocol that will be used for the rest of this article, which can be found in \cite{ishii2010}. Consider the multi-agent system from the previous section. The random manner is specified by the random process $\gamma(k)\in V$ where $k\in \mathbb{Z}_{+}$ is called the time slot. At time slot $k$, $\gamma(k) = i$ indicates that agent $i$ wakes up, then it will choose another neighbor $j$ with a probability ${\rm P}_{ij}$ to communicate, and both agents update their state values. $\gamma(k)$ is assumed to be i.i.d., and its probability distribution is given by
\begin{equation}\label{manner}
{\rm P}(\gamma(k)=i)=\frac{1}{n}, \forall k\in \mathbb{Z}_{+}.
\end{equation}
This means the distribution of agents choosing to wake up at any time slot is uniform. This protocol can be implemented distributively by giving each agent an independent clock that ``ticks'' (wakes the agent up) at the times of an identical stochastic process. For example, \cite{boyd2006} sets a clock that
``ticks'' are distributed as a rate 1 Poisson process for each agent. In this paper, we do not restrict to any particular stochastic process for the purpose of simplicity. Because time slots are the only instances when the value of each agent changes \cite{boyd2006}, we will utilize them as a new time axis in all of our results. The matrix weighted consensus laws for both leaderless and leader-following topologies based on the discussed protocol will be introduced in the next section.

{black}{
\begin{remark}
    In this paper, the assumption of uniform clock distribution is adapted from that in \cite{boyd2006, 4787122}. In the case that the clock distribution setting is nonuniform, i.e.,
\begin{equation}\label{nonuniform}
{\rm P}(\gamma(k)=i) = p_i > 0, \forall k\in \mathbb{Z}_{+},
\end{equation}
where $\sum_{i \in V} p_i = 1$, the definition of expected matrix weight is changed to $\mathbf{M}_{ij}=p_i{\rm P}_{ij}\mathbf{A}_{ij}+p_j{\rm P}_{ji}\mathbf{A}_{ji}$. Thus, all of our results obtained in this paper will remain unchanged. {Additionally, the definition of $\mathbf{M}_{ij}$ reveals its physical meaning. Suppose that the matrix weight $\mathbf{A}_{ij}$ encodes the directional relationship from $i$ to $j$. As it is clearly seen, the probability of agent $i$ to wake up and select $j$ for gossiping is $p_i{\rm P}_{ij}$ and conversely, $p_i{\rm P}_{ji}$ represents the probability of the converse  direction. Thus, $\mathbf{M}_{ij}$, the expected interaction or influence between agent $i$ and $j$, reflects a bidirectional relationship between two agents.}
\end{remark}
}

\begin{lemma}
\emph{(Markov inequality)} If a random variable X can only take non-negative values, then \cite{prob}
   $${\rm P}(X>a) \leq \frac{{{\rm E}}[X]}{a}, \ \  \forall a >0,$$
where ${\rm E}[X]$ is the expectation of X.
\end{lemma}
\section{Randomized matrix weighted consensus with Leaderless topology}
\label{sec:leaderless}
In this section, we propose the randomized matrix weighted consensus law with leaderless topology. Consider a system consisting of $n$ agents whose interconnections between agents are $\mathbf{A}_{ij}\in\mathbb{R}^{d\times d}$. Each agent $i\in V$ has a state vector $\overline{x}_i \in \mathbb{R}^d$. When an agent $i$ wakes up at the $k^{\text{th}}$ time slot, it contacts a neighbor $j$ with a probability ${\rm P}_{ij}$ and both agents update their current state vectors according to
\begin{equation}\label{algorithm1}
    \begin{aligned}
    \overline{x}_i(k+1)&=\overline{x}_i(k)-\alpha_i \mathbf{A}_{ij}\big(\overline{x}_i(k)-\overline{x}_j(k)\big),\\
    \overline{x}_j(k+1)&=\overline{x}_j(k)-\alpha_j \mathbf{A}_{ij}\big(\overline{x}_j(k)-\overline{x}_i(k)\big),
    \end{aligned}
\end{equation}
where $\alpha_i > 0$ is updating step size of agent $i$, and will be designed later. {black}{Meanwhile, the other $n-2$ agents in the network do not involve in this update instant and their state vectors remain unchanged
\begin{equation}
    \begin{aligned}
    \overline{x}_l(k+1)=\overline{x}_l(k), \ \ \forall l \notin \{i,j \} .
    \end{aligned}
\end{equation}
}
\begin{assume}\label{pii}
When an agent wakes up, it {has to choose} another agent in the network to exchange information, that is, ${\rm P}_{ii}=~ 0$ for all $i\in V$.
\end{assume}
{Denote $\overline{\mathbf{x}}(k) = \text{vec}(\overline{{x}}_i(k))$, where $\text{vec}(\cdot)$ is the vectorization operator. Thus, (\ref{algorithm1}) can be rewritten as follow:}
\begin{equation}\label{algorithm11}
    \overline{\mathbf{x}}(k+1)=W_{ij}\overline{\mathbf{x}}(k),
\end{equation}
where $W_{ij}$ is a block matrix
{\small
\begin{equation*}
W_{ij}=\begin{bmatrix}
    &\mathbf{I}_d        &\cdots     &\mathbf{0}           &\cdots    &\mathbf{0} &\cdots &\mathbf{0} \\
    &\vdots     &\ddots     &\vdots          &\ddots    &\vdots &\ddots &\vdots \\
    &\mathbf{0}       &\cdots  &\mathbf{I}_d-\alpha_i \mathbf{A}_{ij} &\cdots    &\alpha_i \mathbf{A}_{ij} &\cdots &\mathbf{0} \\
    &\vdots     &\ddots     &\vdots          &\ddots    &\vdots &\ddots &\vdots \\
    &\mathbf{0}      &\cdots  &\alpha_j \mathbf{A}_{ij} &\cdots    &\mathbf{I}_d-\alpha_j\mathbf{A}_{ij} &\cdots &\mathbf{0}  \\
    &\vdots     &\ddots     &\vdots          &\ddots    &\vdots &\ddots &\vdots \\
    &\mathbf{0}       &\cdots     &\mathbf{0}            &\cdots    &\mathbf{0} &\cdots &\mathbf{I}_d
\end{bmatrix}
\end{equation*}}
in which $\mathbf{I}_d$ denotes the $d\times d$ identity matrix, $\mathbf{0}$ denotes the $d\times d$ zero matrix, $\mathbf{I}_d-\alpha_i \mathbf{A}_{ij}$ is the block entry of matrix $W_{ij}$ in the $({(i-1)d+1:id})^{\text{th}}$ rows
and $({(i-1)d+1:id})^{\text{th}}$ columns. Block $\mathbf{I}_d-\alpha_j \mathbf{A}_{ij}$ is in the $({(j-1)d+1:jd})^{\text{th}}$ rows
and $({(j-1)d+1:jd})^{\text{th}}$ columns of $W_{ij}$. Due to the symmetry of $\mathbf{A}_{ij}$, $W_{ij}$ is also symmetric.

At a random $k^{\text{th}}$ time slot, we can write
\begin{equation}\label{algorithm13}
\overline{\mathbf{x}}(k+1)=W(k)\overline{\mathbf{x}}(k),
\end{equation}
where the random variable $W(k)$ is drawn i.i.d from some distribution on the set of possible values $W_{ij}$ \cite{boyd2006}. Note that the probability that $W(k)$ takes a specific value $W_{ij}$ is $\frac{1}{n}{\rm P}_{ij}$ (the probability that agent $i$ will wake up at $k^{\text{th}}$ time slot is $\frac{1}{n}$, and the probability that agent $j$ will be chosen by $i$ is ${\rm P}_{ij}$). The expectation of $W(k)$ is thus obtained as
\begin{equation}
    \begin{aligned}
        \overline{W}={\rm E}[W]=\sum_{i \in V}\sum_{j \in V} \frac{1}{n}{\rm P}_{ij} W_{ij}.
    \end{aligned}
\end{equation}
The block entries of $\overline{W}$ are determined as follow
\begin{itemize}
    \item If $i=j$, then
    \begin{align*}
        \overline{W}&(i,i)=\bigg(1-\frac{1}{n}\sum_{j\in V}({\rm P}_{ij}+{\rm P}_{ji})\bigg)\mathbf{I}_d \nonumber\\
        &+\frac{1}{n}\sum_{j \in V}\big({\rm P}_{ij}(\mathbf{I}_d-\alpha_i \mathbf{A}_{ij})+{\rm P}_{ji}(\mathbf{I}_d-\alpha_i \mathbf{A}_{ji})\big) \nonumber\\
        &\qquad= \mathbf{I}_d - \alpha_i\frac{1}{n}\sum_{j \in V}({\rm P}_{ij} \mathbf{A}_{ij}+{\rm P}_{ji} \mathbf{A}_{ji}) \nonumber\\
        &\qquad= \mathbf{I}_d-\alpha_i\sum_{j \in V}\mathbf{M}_{ij}.
    \end{align*}
    \item If $i \neq j$, then
\begin{align*}
\overline{W}(i,j)&=\frac{\alpha_i}{n} 
({\rm P}_{ij} \mathbf{A}_{ij} + {\rm P}_{ji} \mathbf{A}_{ji}) =\alpha_i 
\mathbf{M}_{ij}.
\end{align*}
\end{itemize}
As a result, {the updating algorithm} \eqref{algorithm13} can be rewritten as
\begin{equation}\label{wbar}
\begin{aligned}
\overline{W}&=\mathbf{I}_{dn} - \mathbf{G}(\mathbf{D}^{\text{M}}-\mathbf{A}^\text{M})=\mathbf{I}_{dn} - \mathbf{G}\mathbf{L}^{\text{M}},
\end{aligned}
\end{equation}
where $\mathbf{G}=\text{blkdiag}\left\{\alpha_1\mathbf{I}_d,\dots,\alpha_n\mathbf{I}_d\right\}$.
\subsection{Convergence in expectation}
The solution of (\ref{algorithm13}) can be easily obtained as:
\begin{equation}\label{sol}
\overline{\mathbf{x}}(k+1) = \underbrace{W(k)W(k-1)\dots W(0)}_{:=\phi(k)} \overline{\mathbf{x}}(0).
\end{equation}
In order to prove the convergence in expectation, we compute the mean of $\phi(k)$
\begin{equation}\label{mean}
    \begin{aligned}
        {\rm E}[\phi(k)]&={\rm E}[W(k)W(k-1)\dots W(0)] \\
        &={\rm E}[W(k)]{\rm E}[W(k-1)]\dots{\rm E}[W(0)]\\
        &=\overline{W}^{k+1}=\big(\mathbf{I}_{dn}-\mathbf{G}(\mathbf{D}^\text{M}-\mathbf{A}^\text{M})\big)^{k+1}.
    \end{aligned}
\end{equation}
The spectral properties of $\overline{W}$ are stated in the following lemma, whose proof can be found in \cite{quoc2021}.
\begin{lemma}\label{stepsize}
Let the step sizes satisfy $\alpha_i < \frac{1}{\lVert \mathbf{D}^{{\rm M}}_{i} \rVert}, ~\forall i\in V$. The following statements hold for the matrix $\overline{W}= \mathbf{I}_{dn}-\mathbf{G}(\mathbf{D}^{\rm M}-\mathbf{A}^{\rm M})$.
\begin{itemize}
    \item All eigenvalues of $\overline{W}$ are contained in $(-1,1]$ and the spectral radius is $\rho(\overline{W})=1$ with the corresponding eigenvectors $\mathbf{\textit{v}}\in {\rm null}(\mathbf{L}^{\rm M})$.
    \item The unity eigenvalue 1 of $\overline{W}$ is semi-simple, and $\overline{W}^{\infty}:=\lim_{k\to \infty} \overline{W}^{k}$ exists and is finite. 
\end{itemize}
\end{lemma}
Regarding the existence of $\overline{W}^{\infty}$, we have 
\begin{equation}
    \begin{aligned}
        \overline{W}^{\infty}&=(\mathbf{V}\mathbf{J}\mathbf{V}^{-1})^{\infty}=\mathbf{VJ^{\infty}V^{-1}}\\
        &=\mathbf{V}\text{blkdiag}(1,\dots,1,\mathbf{J}_{l_2}^{\infty},\dots,\mathbf{J}_{l_p}^{\infty})\mathbf{V}^{-1}\\
        &=\mathbf{V}\text{blkdiag}(1,\dots,1,0,\dots,0)\mathbf{V}^{-1}=\sum_{i=1}^{l_1}v_i u_i^{\top},
    \end{aligned}
\end{equation}
where $\mathbf{V}=[v_1,\dots,v_{dn}]$ and $\mathbf{V}^{-1}=[u_1,\dots,u_{dn}]^{\top}$ contain the left- and right eigenvectors of $\overline{W}$, respectively. In the Jordan form 
$\mathbf{J}=\text{blkdiag}(1,\dots,1,\mathbf{J}_{l_2},\dots,\mathbf{J}_{l_p}) \in \mathbb{R}^{dn\times dn}$, the Jordan block $\mathbf{J}_{l_i},i=2,\ldots,p,$ corresponds to the eigenvalue $\lambda_i$ with magnitude strictly smaller than 1. It follows that
\begin{align}
    \lim_{k\to \infty} {\rm E}[\bar{\mathbf{x}}(k)]= \left( \lim_{k\to \infty} \bar{W}^k \right) \bar{\mathbf{x}}(0) = \sum_{i=1}^{l_1}v_i u_i^{\top} \bar{\mathbf{x}}(0).
\end{align}
Combining with Lemma~\ref{nullspace}, we have the following theorem.

\begin{theorem}\label{conthm}
Select the updating step-sizes to satisfy Lemma~\ref{stepsize}. For all $\mathbf{\overline{x}}(0)$, the solution $\mathbf{\overline{x}}(k)$ of \eqref{sol} converges in the expectation to $\mathbf{\overline{x}}^{*}=\mathbf{1}_n\otimes \big([{{u}}_1, {{u}}_2,\dots,{{u}}_d]^\top \mathbf{x}(0) \big)$, if and only if {\rm null}$(\mathbf{L}^{\rm M})=${\rm range}$(\mathbf{1}_n \otimes \mathbf{I}_d)$.
\end{theorem}
\begin{remark} \cite{THM2018} The existence of a positive spanning tree in $\mathcal{G}^{\rm M}$ is a sufficient condition for ${\rm null}(\mathbf{L}^{\rm M})={\rm range}(\mathbf{1}_n\otimes \mathbf{I}_d)$.
 \end{remark}
 \subsection{Randomized matrix weighted average consensus}
 When all agents adopt the same updating step size $\alpha_i = \alpha,~\forall i\in V$, the matrix $\overline{W}$ becomes symmetric, and $\mathbf{V}^{\top}=\mathbf{V}^{-1}$. The following result follows from Theorem~\ref{conthm}.
 \begin{theorem}\label{avconthm}
Suppose that all agents use the same step-size $\alpha < \frac{1}{\max_{i\in V}\lVert \mathbf{D}^{\rm M}_{i} \rVert}$. For all $\mathbf{\overline{x}}(0)$, the solution $\mathbf{\overline{x}}(k)$ of (\ref{sol}) converges in expectation to the average vector $\mathbf{\overline{x}}^{*}= \mathbf{1}_n \otimes \hat{\overline{\mathbf{x}}}$, where $\hat{\overline{\mathbf{x}}}=\frac{1}{n}\big(\mathbf{1}_n^\top\otimes \mathbf{I}_d\big)\mathbf{\overline{x}}(0)$, if and only if {\rm null}$(\mathbf{L}^{\rm M})={\rm range}(\mathbf{1}_n\otimes \mathbf{I}_d)$.
\end{theorem}

Although we have shown that, if the graph $\mathcal{G}$ satisfies a mild assumption and the step size is  small enough, it is guaranteed that the agents reach the average consensus in expectation. However, we have not yet mentioned the convergence rate of these agents. The following section  will describe a method to quantify the average convergence rate of agents. It will be shown that the convergence rate of the agents is closely related to the second-largest amplitude eigenvalue of the matrix ${\rm E}[W(k)^\top W(k)]$. Inspired by \cite{boyd2006}, we first introduce our quantity of interest
\begin{define}
 \emph{($\epsilon$-consensus time)} For any $0<\epsilon<1$, the $\epsilon$-consensus time is defined as
\begin{equation*}
    T(\epsilon)=\underset{\overline{\mathbf{x}}(0)}{\sup}\inf \bigg( k:{\rm P}\bigg(\frac{\lVert \mathbf{\overline{x}}(k)-\mathbf{1}_n\otimes \hat{\mathbf{\overline{x}}} \rVert}{\lVert \mathbf{\overline{x}}(0)-\mathbf{1}_n\otimes \hat{\mathbf{\overline{x}}} \rVert} \geq \epsilon\bigg)\leq \epsilon\bigg).
\end{equation*}
\end{define}
Intuitively, $T({\epsilon})$ represents the number of clock ticks needed for the trajectory $\mathbf{\overline{x}}$ to reach the consensus value with a high probability. It is worth noting that $\epsilon$ measures both accuracy and success probability and is often set to $\frac{1}{n}$ \cite{TIT2010}. In this paper, we provide the upper bound formula for the proposed average consensus algorithm. Firstly, the convergence of the second moment will be proven.
\subsubsection{Convergence of the second moment}
Define the error vector 
$\overline{y}_i(k)=\overline{x}_i(k)-\hat{\overline{\mathbf{x}}}$, and
$\mathbf{\overline{y}}(k)=\text{vec}(\overline{y}_i)=\mathbf{\overline{x}}(k)-\mathbf{1}_n\otimes \hat{\mathbf{\overline{x}}}$, we have
\begin{equation}\label{y}
\begin{aligned}
    \mathbf{\overline{y}}(k+1)&= \mathbf{\overline{x}}(k+1)-\mathbf{1}_n\otimes \hat{\mathbf{\overline{x}}}\\
    &=W(k)\mathbf{\overline{x}}(k)-W(k)(\mathbf{1}_n\otimes \mathbf{I}_d)\hat{\mathbf{\overline{x}}}\\  
    &=W(k)\mathbf{\overline{y}}(k).
\end{aligned}
\end{equation}
Thus, $\mathbf{\overline{y}}(k)$ has the same linear system as $\mathbf{\overline{x}}(k)$. From \eqref{y}, we have the following equation \cite{boyd2006}
\begin{equation}\label{minh}
    \begin{aligned}
        {\rm E}[\mathbf{\overline{y}}(k+1)^{\top}\mathbf{\overline{y}}(k+1)|\mathbf{\overline{y}}(k)]
        = \mathbf{\overline{y}}(k)^{\top}{\rm E}[W(k)^{\top}W(k)]\mathbf{\overline{y}}(k).
    \end{aligned}
\end{equation}
Similar to $W(k)$, we can consider $W(k)^\top W(k)$ as a random variable which is drawn i.i.d from some distribution on the set of possible values $W_{ij}^\top W_{ij}$. Some properties of ${\rm E}[W(k)^\top W(k)]$ are provided by the following lemmas.
 \begin{lemma}\label{stepsizeav}
Let the updating step-size of each agent satisfy $\alpha <{\min}(\frac{1}{\max_{i\in V}\lVert \mathbf{D}^{\rm M}_{i} \rVert},\frac{1}{\max_{i,j\in V}\lVert \mathbf{A}_{ij} \rVert})$, the following statements hold
\begin{enumerate}
    \item The set of all real symmetric matrices with eigenvalues in $[0,1]$ is convex. Every possible matrix $W_{ij}^\top W_{ij}$ is in this set.
    \item ${\rm E}[W(k)^\top W(k)]$ has a unity spectral radius, and its unity eigenvalues are semi-simple.
    \item $\mathbf{1}_n \otimes\mathbf{I}_d$ are 
only $d$ orthogonal right eigenvectors corresponding to the unity eigenvalue $\lambda=1$ of ${\rm E}[W(k)^\top W(k)]$ if and only if {\rm null}$(\mathbf{L}^{\rm M})= {\rm range}(\mathbf{1}_n\otimes \mathbf{I}_d)$.
\end{enumerate}
    
\emph{Proof:} See Appendix A.
\end{lemma}
Under the assumption null$(\mathbf{L}^{\text{M}})=\text{range}(\mathbf{1}_n\otimes \mathbf{I}_d)$ on our hand, $[v_1,\dots,v_d] =\mathbf{1}_n \otimes\mathbf{I}_d$ are 
only $d$ orthogonal right eigenvectors corresponding to the unity eigenvalue $\lambda=1$ of ${\rm E}[W(k)^\top W(k)]$. By reason of $\mathbf{\overline{y}}(k)\perp \text{span}\big( v_1,\dots,v_d \big)$, the Rayleigh-Ritz theorem states that
\begin{align} \label{raritz}
\mathbf{\overline{y}}(k)^\top &{\rm E}[W(k)^\top W(k)] \mathbf{\overline{y}}(k) \nonumber\\
      &\leq \lambda_{d+1}({\rm E}[W(k)^\top W(k)]) \mathbf{\overline{y}}(k)^\top \mathbf{\overline{y}}(k),
\end{align}
where $\lambda_{d+1}<1$ is the second largest eigenvalue of ${\rm E}[W(k)^\top W(k)]$. After iterating (\ref{minh}) and (\ref{raritz}), we get:
\begin{equation}\label{raritz2}
    \begin{aligned}
      {\rm E}[\mathbf{\overline{y}}(k)^\top \mathbf{\overline{y}}(k)] \leq \lambda_{d+1}^{k}({\rm E}[W(k)^\top W(k)]) \mathbf{\overline{y}}(0)^\top \mathbf{\overline{y}}(0),
    \end{aligned}
\end{equation}
which also implies that the convergence of the second moment of the proposed algorithm. 

When the matrix weight $\mathbf{A}_{ij}$ is nonsymmetric in general, necessary and sufficient conditions can be obtained by applying the results proposed in \cite{boyd2006,1272421}, which are as follows
\begin{lemma}\label{iff}
 {The first and second moment of the solution of (\ref{algorithm11}) will converge in probability if and only if there exists a common stepsize $\alpha$ such that
\begin{enumerate}
    \item $\rho\big(\overline{W}-\frac{(\mathbf{1}_n\mathbf{1}_n^\top)\otimes\mathbf{I}_d}{n}\big)<1$,
    \item $\rho\big({\rm E}[W(k)\otimes W(k)]-\frac{(\mathbf{1}_{n^2}\mathbf{1}_{n^2}^\top)\otimes\mathbf{I}_{d^2}}{n^2}\big)<1$.
\end{enumerate}
{black}{\emph{Proof:} See Appendix B.}
}
\end{lemma}
{In Lemma \ref{iff}, the first constraint makes all agents reach the average consensus, whereas the second guarantees the convergence of the second moment. However, despite having explicit criteria for the convergence of the second moment, it is clearly seen that these requirements are not intuitive and quite hard to assess. Moreover, the symmetry of $\mathbf{A}_{ij}$ allows us to find interesting results of $\epsilon$-consensus time, which will be provided later.}
\subsubsection{Upper bound of the $\epsilon$-consensus time}
\begin{theorem}\label{upperbound1}
    Select a common step size for every agent such that $\alpha < {\min}(\frac{1}{\max_{i\in V}\| \mathbf{D}^{\rm M}_{i} \|},\frac{1}{\max_{i,j \in V} \| \mathbf{A}_{ij} \|})$. With an arbitrary initial  state vector $\mathbf{\overline{x}}(0)$, the solution $\mathbf{\overline{x}}(k)$ of \eqref{sol} converges in the expectation to the average vector $\mathbf{\overline{x}}^{*}=\bigg(\frac{1}{n} \mathbf{1}_n\mathbf{1}_n^\top \otimes \mathbf{I}_d \bigg)\mathbf{\overline{x}}(0)$ if and only if ${\rm null}(\mathbf{L}^{\rm M})={\rm range}(\mathbf{1}_n\otimes \mathbf{I}_d)$. Furthermore, the $\epsilon$-consensus time is upper bounded by a function of the second largest eigenvalue of ${\rm E}[W(k)^\top W(k)])$.
\end{theorem}
\emph{Proof:} Using Markov's inequality, we have:
\begin{align}
        {\rm P}\bigg(\frac{\lVert \mathbf{\overline{x}}(k)-\mathbf{1}_n\otimes \hat{\mathbf{\overline{x}}} \rVert}{\lVert \mathbf{\overline{x}}(0)-\mathbf{1}_n\otimes \hat{\mathbf{\overline{x}}} \rVert} & \geq \epsilon \bigg) 
        ={\rm P}\bigg(\frac{\mathbf{\overline{y}}(k)^\top \mathbf{\overline{y}}(k)}{\mathbf{\overline{y}}(0)^\top \mathbf{\overline{y}}(0)}\geq \epsilon^2\bigg) \nonumber\\
        &\leq\frac{\epsilon^{-2}{\rm E}[\mathbf{\overline{y}}(k)^\top \mathbf{\overline{y}}(k)]}{\mathbf{\overline{y}}(0)^\top \mathbf{\overline{y}}(0)} \nonumber\\
        &\leq \epsilon^{-2} \lambda_{d+1}^k \big( {\rm E}[{W}(k)^\top W(k)] \big).
\end{align}

As a result, for $k \geq K(\epsilon)=\frac{3{\log}(\epsilon^{-1})}{{\log}\lambda_{d+1}^{-1} \big( {\rm E}[{{W}}(k)^\top {{W}}(k)] \big)}$, there holds
$$\rm{P}\bigg(\frac{\lVert \mathbf{\overline{x}}(k)-\mathbf{1}_n\otimes \hat{\mathbf{\overline{x}}} \rVert}{\lVert \mathbf{\overline{x}}(0)-\mathbf{1}_n\otimes \hat{\mathbf{\overline{x}}} \rVert} \geq \epsilon \bigg) \leq \epsilon.$$ 
This implies $K(\epsilon)$ is the upper bound of the $\epsilon$-consensus time. 

\section{Randomized matrix weighted consensus with leader-following topology}
\label{sec:leader-follower}
Consider the previous graph $\mathcal{G}$, and then add a node $i=0$ to represent the leader. This leader node has a state vector $\mathbf{\overline{x}}_0$. A set of directed edges $E_0$ from vertex $0$ to some vertices $i \in V$, and a corresponding set of matrix-weights $\mathcal{A}_0=\left\{\mathbf{A}_{i0} = \mathbf{A}_{i0}^\top \geq 0, \forall i\in V\right\}$. $\mathbf{A}_{i0}=0$ represents the situation in which agent $i$ has no connection to the leader. The leader-following system is said to achieve a consensus if, for any initial state $\mathbf{\overline{x}}_i(0)\in\mathbb{R}^d, i\in V$, there holds $\lim_{t\rightarrow\infty} \mathbf{\overline{x}}_i(t)\rightarrow \mathbf{\overline{x}}_0$. In \cite{Nguyen2022MCA}, the authors proposed a continuous deterministic algorithm where the consensus phenomena are completely proven. {In this section, we first propose a discrete-time version of the algorithm proposed in \cite{Nguyen2022MCA}. Then, a randomized version of the algorithm will be introduced.}
\subsection{Discrete-time matrix weighted consensus with Leader-Following topology}
Our deterministic consensus algorithm for the Leader-Following topology is as follows: in particular, each agent $i \in V$ updates its value to $\mathbf{\overline{x}}_0(k+1)$ via
\begin{equation}\label{algorithm2}
    \begin{aligned}
    \overline{x}_0(k+1)&=\overline{x}_0(k)\\
    \overline{x}_i(k+1)&=\overline{x}_i(k)+ \theta\alpha \sum_{j\in \mathcal{N}_i} \mathbf{A}_{ij}\big(\overline{x}_j(k)-\overline{x}_i(k)\big)\\
    &\qquad\quad + (1-\theta)\alpha  \mathbf{A}_{i0}\big(\overline{x}_0 - \overline{x}_i(k)\big),\\
    \end{aligned}
\end{equation}
where $0<\theta<1$ and the step sizes $\alpha > 0$ will be designed later. To ensure the system reaches a consensus, we state the following assumptions:
\begin{assume}\label{Ai0}
    $\sum_{i \in V} \mathbf{A}_{i0} >0$.
\end{assume}
\begin{assume}\label{nullL}
    {\rm null}$(\mathbf{L}) = {\rm range}(\mathbf{1}_n \otimes \mathbf{I}_d)$.
\end{assume}
Under these assumptions, we have the following theorem.
\begin{theorem}\label{ddmwc}
    Select $\alpha<\min(\frac{1}{\max_{i\in V_f}\lVert \mathbf{D}_{i} \rVert}, \frac{2}{ \max_{i\in V_f}(\lambda(\mathbf{A}_{i0}))})$, for any initial condition $\mathbf{\overline{x}}(0)$, the system \eqref{algorithm2} achieves leader-follower consensus at geometric rate.
\end{theorem}
\emph{Proof:} See Appendix C.
\subsection{Randomized matrix weighted consensus with Leader-Following topology}
The randomized version of matrix weighted consensus with Leader-Following topology is thus described as follows: Each agent $i\in V$ stores its own local vector $\overline{x}_i \in \mathbb{R}^d$. When an agent $i$ wakes up at the $k^{\text{th}}$ time slot, it will contact {black}{only one} neighbor $j$ with a probability ${\rm P}_{ij}$ and both agents will update their current state vector as
\begin{equation}\label{algorithm3}
    \begin{aligned}
    \overline{x}_i(k+1)&=\overline{x}_i(k)+ \theta\alpha \mathbf{A}_{ij} \big(\overline{x}_j(k)-\overline{x}_i(k)\big)\\
    &\qquad + (1-\theta)\alpha \mathbf{A}_{i0} \big(\overline{x}_0-\overline{x}_i(k)\big)\\
    \overline{x}_j(k+1)&=\overline{x}_j(k)+ \theta\alpha \mathbf{A}_{ij}\big(\overline{x}_i(k)-\overline{x}_j(k)\big)\\
    &\qquad + (1-\theta)\alpha \mathbf{A}_{j0}\big(\overline{x}_0 - \overline{x}_j(k)\big),\\
    \end{aligned}
\end{equation}
where $0<\theta<1$, $\alpha > 0$ are step sizes. {black}{Meanwhile, the $n-2$ remaining agents in the network will maintain their current state values.}
\begin{define}
\emph{(Expected leader-follower weight)} An expected leader-follower weight of agent $i\in V$ is defined as
$$\mathbf{M}_{i0}=\frac{1}{n} \sum_j({\rm P}_{ij}+{\rm P}_{ji}) \mathbf{A}_{i0}.$$
\end{define}
The following assumptions are adopted:
\begin{assume}\label{Mi0}
    $\sum_{i \in V} \mathbf{M}_{i0} >0$,
\end{assume}
\begin{assume}\label{Lm}
   {\rm null}$(\mathbf{L}^{\rm M})={\rm range}(\mathbf{1}_n\otimes \mathbf{I}_d)$.
\end{assume}
{black}{Similar to leaderless topology, the existence of a positive spanning tree among followers is sufficient for Assumption \ref{Lm} to hold.} Denote the error vectors $\mathbf{\overline{y}}_i(k)=\mathbf{\overline{x}}_i(k)-\mathbf{\overline{x}}_0$ and $\mathbf{\overline{y}}(k)=\text{vec}(\overline{y}_i(k))$. With a probability $\frac{1}{n}{\rm P}_{ij}$, we have
\begin{equation}\label{algorithm31}
    \begin{aligned}
        \overline{\mathbf{y}}(k+1)=W_{ij}\overline{\mathbf{y}}(k),
    \end{aligned}
\end{equation}
where $W_{ij}$ can be expressed as \eqref{algorithm22}. 
\begin{figure*}
    \centering
   {\small
\begin{equation}\label{algorithm22}
    \begin{aligned}
        W_{ij}= \mathbf{I}_{dn}-\alpha\theta
        \begin{bmatrix}
         &\mathbf{0}        &\cdots     &\mathbf{0}             &\cdots    &\mathbf{0} &\cdots &\mathbf{0} \\
        &\vdots     &\ddots     &\vdots          &\ddots    &\vdots &\ddots &\vdots \\
         &\mathbf{0}        &\cdots  &\mathbf{A}_{ij} &\cdots    &-\mathbf{A}_{ij} &\cdots &\mathbf{0}  \\
        &\vdots     &\ddots     &\vdots          &\ddots    &\vdots &\ddots &\vdots \\
        &\mathbf{0}        &\cdots  &- \mathbf{A}_{ij} &\cdots    & \mathbf{A}_{ij} &\cdots &\mathbf{0}  \\
        &\vdots     &\ddots     &\vdots          &\ddots    &\vdots &\ddots &\vdots \\
        &\mathbf{0}        &\cdots     &\mathbf{0}             &\cdots    &\mathbf{0} &\cdots &\mathbf{0}   \\
        \end{bmatrix}
 -\alpha(1-\theta)
    \begin{bmatrix}
         &\mathbf{0}        &\cdots     &\mathbf{0}             &\cdots    &\mathbf{0} &\cdots &\mathbf{0} \\
        &\vdots     &\ddots     &\vdots          &\ddots    &\vdots &\ddots &\vdots \\
         &\mathbf{0}        &\cdots  & \mathbf{A}_{i0} &\cdots    &\mathbf{0} &\cdots &\mathbf{0}  \\
        &\vdots     &\ddots     &\vdots          &\ddots    &\vdots &\ddots &\vdots \\
        &\mathbf{0}        &\cdots  &\mathbf{0} &\cdots    & \mathbf{A}_{j0} &\cdots &\mathbf{0}  \\
        &\vdots     &\ddots     &\vdots          &\ddots    &\vdots &\ddots &\vdots \\
        &\mathbf{0}        &\cdots     &\mathbf{0}             &\cdots    &\mathbf{0} &\cdots &\mathbf{0} ,
    \end{bmatrix}.
    \end{aligned}
\end{equation}}
\end{figure*}
Consider $\overline{W}~={\rm E}[W(k)]=\frac{1}{n}\sum_{i,j} {\rm P}_{ij} W_{ij}$, we have
\begin{align} \label{23}
    \overline{W} &= \mathbf{I}_{dn}- \theta \alpha (\mathbf{D}^{\text{M}}-\mathbf{A}^{\text{M}})-(1-\theta)\alpha \text{blkdiag}(\mathbf{M}_{i0}) \nonumber\\
    &=\mathbf{I}_{dn}- \alpha \big(\theta  \mathbf{L}^{{\text{M}}}+(1-\theta) \text{blkdiag}(\mathbf{M}_{i0})\big).
\end{align}
Under the Assumptions \ref{Mi0} and \ref{Lm}, the following theorem yields
\begin{theorem}\label{sdmwc}
 Let $\alpha<\min(\frac{1}{ \underset{i \in V_f}{\max}\lVert{\mathbf{D}^{\rm M}_{i}\rVert}},\frac{2}{ \underset{i\in V_f}{\max}(\lambda(\mathbf{M}_{i0}))})$, for all initial condition $\mathbf{\overline{x}}(0)$,  system \eqref{algorithm3} exponentially achieves a leader-follower consensus in expectation. 
\end{theorem}
 \emph{Proof:} The proof of this theorem is similar to Appendix B.

 To check the convergence of the second moment and then determine the convergence rate, we again study the spectral properties of ${\rm E}[W(k)^\top W(k)]$.

 \begin{theorem}\label{2mmlf}
Let $\alpha$ satisfy Theorem \ref{sdmwc} and $\alpha<\min(\frac{2}{{\max_{i,j\in V}}\lVert\mathbf{A}_{ij} \rVert},\frac{2}{ \underset{i \in V_f}{\max}(\lambda(\mathbf{A}_{i0}))})$, for any initial condition $\mathbf{\overline{x}}(0)$, the spectral radius of ${\rm E}[W(k)^\top W(k)]$ is strictly less than 1, implying that the proposed algorithm's second moment has converged.
\end{theorem}
\emph{Proof:} See Appendix D.

\subsubsection{Upper bound of the $\epsilon$-consensus time 
}
\begin{theorem}\label{upperbound2}
    Select a common step-size for every agent to satisfy Theorem~\ref{2mmlf}. For all $\mathbf{\overline{x}}(0)$, the solution $\mathbf{\overline{x}}(k)$ of (\ref{algorithm3}) converges in the expectation to the leader's state vector. Furthermore, the $\epsilon$-consensus time is upper bounded by a function of the spectral radius of ${\rm E}[W(k)^\top W(k)])$.
\end{theorem}
\emph{Proof:} We have $\mathbf{\overline{y}}(k)^\top {\rm E}[W(k)^\top W(k)] \mathbf{\overline{y}}(k)$
\begin{align*}
      &\leq \lambda_{\max}({\rm E}[W(k)^\top W(k)]) \mathbf{\overline{y}}(k)^\top \mathbf{\overline{y}}(k)\\
      &\leq \lambda^k_{\max}({\rm E}[W(k)^\top W(k)]) \mathbf{\overline{y}}(0)^\top \mathbf{\overline{y}}(0),
\end{align*}
where {$\lambda_{\max}({\rm E}[W(k)^\top W(k)])<1$} is the maximum eigenvalue of ${\rm E}[W(k)^\top W(k)]$. It follows from the Markov's inequality that
\begin{align*}
        {\rm P}\bigg( &\frac{\lVert \mathbf{\overline{x}}(k)-\mathbf{1}_n\otimes \mathbf{\overline{x}}_0 \rVert}{\lVert \mathbf{\overline{x}}(0)-\mathbf{1}_n\otimes \mathbf{\overline{x}}_0 \rVert}  \geq \epsilon \bigg) 
        ={\rm P}\bigg(\frac{\mathbf{\overline{y}}(k)^\top \mathbf{\overline{y}}(k)}{\mathbf{\overline{y}}(0)^\top \mathbf{\overline{y}}(0)}\geq \epsilon^2\bigg) \nonumber\\
        &\leq\frac{\epsilon^{-2}{\rm E}[\mathbf{\overline{y}}(k)^\top \mathbf{\overline{y}}(k)]}{\mathbf{\overline{y}}(0)^\top \mathbf{\overline{y}}(0)} \leq \epsilon^{-2} \lambda_{\max}^k \big( {\rm E}[W(k)^\top W(k)] \big).
\end{align*}
As a result, for $k \geq K(\epsilon)=\frac{3{\log}(\epsilon^{-1})}{{\log}\lambda_{max}^{-1} \big( {\rm E}[W(k)^\top W(k)] \big)}$, there holds
$\rm{P}\bigg(\frac{\lVert \mathbf{\overline{x}}(k)-\mathbf{1}_n\otimes \mathbf{\overline{x}}_0 \rVert}{\lVert \mathbf{\overline{x}}(0)-\mathbf{1}_n\otimes \mathbf{\overline{x}}_0 \rVert} \geq \epsilon \bigg) \leq \epsilon.$ 
Thus $K(\epsilon)$ is the upper bound of the $\epsilon$-consensus time. 
\section{Application in bearing-based network localization problem}\label{section: NL}

Due to its importance in network operations and several application tasks, distributed localization of sensor networks has received significant research attention. Bearing-based network localization, which refers to a class of algorithms where the network configuration is specified by the bearing (direction/line-of-sight) vector between them, was proposed in \cite{bearingzhao}. In these types of algorithms, all agents just have to account for the minimum amount of bearing sensing capacity compared to distance and position-based localization methods, thus, reducing the deployment cost. In this section, we extend the idea of randomized matrix-weight consensus to present a randomized network localization algorithm. {For an in-depth exploration of the content presented in this section, we encourage readers to refer to our conference paper \cite{LPNM2023ICCAIS}.}

\subsection{Problem Formulation}

A sensor network of $n$ nodes can be considered a multi-agent system where each node (or agent) $i \in \left\{ 1,2,...,n\right\}$ has an absolute position $\overline{x}_i \in \mathbb{R}^d$ (which needs to be estimated). 
Suppose that $\overline{x}_i \neq \overline{x}_j$, the \emph{bearing vector} between two agents $i$ and $j$ is defined as \cite{bearingzhao}
\begin{equation}\label{bear_def}
    g_{ij}=\frac{\overline{x}_j-\overline{x}_i}{\|\overline{x}_j-\overline{x}_i\|}.
\end{equation}
It can be checked that $\|g_{ij}\|=1$ as $g_{ij}$ is a unit vector. Let the local coordinate systems of all agents in the system be aligned. Then, we have $g_{ij}=-g_{ji}, \forall i, j \in V,~i\neq j$.

{The sensor network can be jointly described by $\mathcal{G}(\bar{\mathbf{x}})$, where $\mathcal{G}=(V,E,A)$ is a matrix weighted graph and  
$\overline{\mathbf{x}}=\text{vec}(\overline{x}_i)$ is called a configuration. The matrix weights in $A$ are orthogonal projection matrices  
\begin{equation}
    \mathbf{A}_{ij}=\mathbf{I}_d-{g_{ij}g_{ij}^{\top}},
\end{equation}
which satisfy $\mathbf{A}_{ij}=\mathbf{A}_{ji}=\mathbf{A}_{ij}^{\top}$. Furthermore, every $\mathbf{A}_{ij}$ is idempotent and positive semidefinite, i.e., $\mathbf{A}_{ij}^2=\mathbf{A}_{ij}\geq 0$, we also have $\text{null}(\mathbf{A}_{ij}) = \text{span}(g_{ij})$ \cite{bearingzhao}. The configuration $\overline{\mathbf{x}}$ is a stacked vector of all sensors nodes' positions. An important question to be addressed is whether or not the network configuration specified by the set of bearing vectors $\{g_{ij}|~(i,j) \in {E}\}$ can be uniquely determined up to a translation and a scaling. A sufficient condition to validate this property is based on the notion of infinitesimal bearing rigidity, which is given as follows.}

\begin{define} \cite{bearingzhao}
 A framework $\mathcal{G}(\overline{\mathbf{x}})$ is infinitesimally bearing rigid if and only if the matrix weighted Laplacian satisfies rank$(\mathbf{L})=dn-d-1$.  
\end{define}

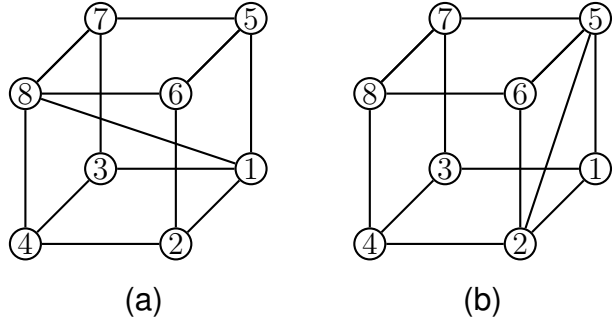
\begin{figure}
\centering
\subfloat[]{
\begin{tikzpicture}[
roundnode/.style={circle, draw=black, thick, minimum size=2mm,inner sep= 0.25mm},
squarednode/.style={rectangle, draw=red!60, fill=red!5, very thick, minimum size=5mm},
]
    \node[roundnode]   (v1)   at   (1,1) {$1$};
    \node[roundnode]   (v2)   at   (0,0) {$2$};
    \node[roundnode]   (v3)   at   (-1,1) {$3$};
    \node[roundnode]   (v4)   at   (-2,0) {$4$};
    \node[roundnode]   (v5)   at   (1,3) {$5$};
    \node[roundnode]   (v6)   at   (0,2) {$6$};
    \node[roundnode]   (v7)   at   (-1,3) {$7$};
    \node[roundnode]   (v8)   at   (-2,2) {$8$};
    \draw[-,thick] (v1)--(v2)--(v4)--(v3)--(v1);
    \draw[-,thick] (v5)--(v6)--(v8)--(v7)--(v5);
    \draw[-,thick] (v1)--(v5)--(v6)--(v2);
    \draw[-,thick] (v3)--(v7)--(v8)--(v4);
    \draw[-,thick] (v1)--(v8);

\end{tikzpicture} \label{rigid example}
}
\qquad
\subfloat[]{ 
\begin{tikzpicture}[
roundnode/.style={circle, draw=black, thick, minimum size=2mm,inner sep= 0.25mm},
squarednode/.style={rectangle, draw=red!60, fill=red!5, very thick, minimum size=5mm},
]
    \node[roundnode]   (v1)   at   (1,1) {$1$};
    \node[roundnode]   (v2)   at   (0,0) {$2$};
    \node[roundnode]   (v3)   at   (-1,1) {$3$};
    \node[roundnode]   (v4)   at   (-2,0) {$4$};
    \node[roundnode]   (v5)   at   (1,3) {$5$};
    \node[roundnode]   (v6)   at   (0,2) {$6$};
    \node[roundnode]   (v7)   at   (-1,3) {$7$};
    \node[roundnode]   (v8)   at   (-2,2) {$8$};
    \draw[-,thick] (v1)--(v2)--(v4)--(v3)--(v1);
    \draw[-,thick] (v5)--(v6)--(v8)--(v7)--(v5);
    \draw[-,thick] (v1)--(v5)--(v6)--(v2);
    \draw[-,thick] (v3)--(v7)--(v8)--(v4);
    \draw[-,thick] (v2)--(v5);
\end{tikzpicture}
\label{norigid example}}
    \caption{Examples of infinitesimally/non-infinitesimally bearing rigid frameworks in three-dimensional space. 
    }
\end{figure}

An example of infinitesimally/non-infinitesimally bearing rigid frameworks in the two-dimensional space is depicted in Fig.~\ref{rigid example}-\ref{norigid example}. Suppose that there are $n_a$ agents ($0\leq n_a \leq n$), known as beacons, who can measure their own real positions. The rest $n_f=n-n_a$ agents are
called followers (Note that the network cannot be localized without beacons). Without loss of generality, we denote the first $n_a$ agents as beacons ($V_a = \left\{1,2,...,n_a\right\}$) and the rest as followers ($V_f = \left\{n_a+1,n_a+2,...,n\right\}$). The bearing-based network localization problem can be stated as follows.

\textbf{Problem 1.} \emph{Assuming framework $\mathcal{G}(\overline{\mathbf{x}})$ is infinitesimally bearing rigid and suppose that there exist at least $2\leq n_a< n$ beacon nodes which know their absolute positions. The initial position estimation of the {network} is $\hat{\overline{x}}(0)$. Design the update law for each agent $\overline{u}_i(k)=\hat{\overline{x}}(k+1)-\hat{\overline{x}}(k)$ $\forall i \in V$ based on the relative estimates $\left\{\hat{\overline{x}}_i(k)-\hat{\overline{x}}_j(k)\right\}$ and the constant bearing measurements $\left\{g_{ij}\right\}$ such that $\hat{\overline{\mathbf{x}}}(k)\rightarrow \overline{\mathbf{x}}$ as $k\rightarrow \infty$ for all $i\in V$.} 

\subsection{Randomized bearing-based network localization}
Consider a network consisting of $n$ sensors (agents). At time slot $k$, suppose that agent $i$ wakes up, then it will choose only one neighbor $j$ with a probability ${\rm P}_{ij}$ to communicate. If both the waken and chosen ones are beacons, then they just retain their values. If both the waken and chosen ones are followers, they will update their values simultaneously. If one of the two agents is a beacon and the other is a follower, only the follower can update its value. In summary, the updating law is designed as follows 

\begin{enumerate}
\item if $i$ and $j$ are beacons:
\begin{equation}\label{algorithmNL0}
    \begin{aligned}
    \overline{x}_i(k+1)&=\overline{x}_i(k)=x_i,\\
    \overline{x}_j(k+1)&=\overline{x}_j(k)=x_j.
    \end{aligned}
\end{equation}

\item if $i$ and $j$ are followers:
\begin{equation}\label{algorithmNL1}
    \begin{aligned}
    \hat{\overline{x}}_i(k+1)&=\hat{\overline{x}}_i(k)-\alpha \mathbf{A}_{ij}\big(\hat{\overline{x}}_i(k)-\hat{\overline{x}}_j(k)\big),\\
    \hat{\overline{x}}_j(k+1)&=\hat{\overline{x}}_j(k)-\alpha \mathbf{A}_{ji}\big(\hat{\overline{x}}_j(k)-\hat{\overline{x}}_i(k)\big).\\
    \end{aligned}
\end{equation}
\item if one of the partners is a beacon and the other is a follower (without loss of generality, assume $i$ is a follower)
\begin{equation}\label{algorithmNL2}
    \begin{aligned}
    \hat{\overline{x}}_i(k+1)&=\hat{\overline{x}}_i(k)-\alpha \mathbf{A}_{ij}\big(\hat{\overline{x}}_i(k)-\overline{x}_j(k)\big),\\
    \overline{x}_j(k+1)&=\overline{x}_j(k)=\overline{x}_j,
    \end{aligned}
\end{equation}
\end{enumerate}
where $\alpha > 0$ is updating step size. {black}{The values of the remaining agents, who are not taking part in the update, are kept unchanged.}


Denote $\Tilde{\overline{x}}_i(k)=\hat{\overline{x}}_i(k)-\overline{x}_i$ $\forall i \in V_f$ and $\Tilde{\overline{\mathbf{x}}} _f=[\Tilde{\overline{x}}_{n_a+1}^{\top},\Tilde{\overline{x}}_{n_a+2}^{\top},...,\Tilde{\overline{x}}_{n}^{\top}]^{\top}$.
We subtract both sides of (\ref{algorithmNL1}) and (\ref{algorithmNL2}) by $p_f$. In addition, due to the fact that $\overline{x}_i-\overline{x}_j \in \text{null} (\mathbf{A}_{ij})$, a quantity $\alpha\mathbf{A}_{ij}(\overline{x}_i-\overline{x}_j)$ is added to the right-hand side of every follower's equation of (\ref{algorithmNL1}) and (\ref{algorithmNL2}). Thus, we can rewrite (\ref{algorithmNL0})-(\ref{algorithmNL2}) as
\begin{equation}
\Tilde{\overline{\mathbf{x}}}_f =W_{ij}\Tilde{\overline{\mathbf{x}}}_f (k),
\end{equation}
where $W_{ij} \in \mathbb{R}^{n_fd\times n_fd}$ and can be determined as
\begin{enumerate}
\item if $i$ and $j$ are beacons:
\begin{equation}\label{algorithm01}
    \begin{aligned}
    W_{ij} &= \mathbf{I}_{dn\times dn}.
    \end{aligned}
\end{equation}

\item if $i$ and $j$ are followers, the updating matrix $W_{ij}$ is as given in \eqref{algorithm11}, with $\alpha_i=\alpha_j=\alpha$. 
\item if one agent $i$ is a follower and the other agent $j$ is a beacon, we have the updating matrix 
\begin{equation}\label{algorithm21}
\begin{aligned}
W_{ij} = \text{blkdiag}(\mathbf{I}_d,\dots,\mathbf{I}_d-\alpha \mathbf{A}_{ij},\dots,\mathbf{I}_d).
\end{aligned}
\end{equation}
\end{enumerate}
It can be seen that for all three scenarios, $W_{ij}$ is symmetric due to the symmetry of $\mathbf{A}_{ij}$. At a random $k^{\text{th}}$ time slot, we now can write
\begin{equation}\label{algorithmNL13}
\Tilde{\overline{\mathbf{x}}}_f (k+1)=W(k)\Tilde{\overline{\mathbf{x}}}_f(k),
\end{equation}
where the random variable $W(k)$ is drawn i.i.d from some distribution on the set of all possible $W_{ij}$. 


\begin{assume}\label{probability}
For every $(i,j)\in E$, ${\rm P}_{ij}+{\rm P}_{ji}>0$.   
\end{assume}
Note that $\mathbf{M}_{ij}=\frac{1}{n}(\mathbf{A}_{ij} {\rm P}_{ij}+\mathbf{A}_{ji} {\rm P}_{ji})=\frac{1}{n}({\rm P}_{ij}+{\rm P}_{ji})\mathbf{A}_{ji}$. Thus, Assumption~\ref{probability} implies that $\mathbf{M}_{ij}$ is positive semi-definite if and only if $\mathbf{A}_{ij}$ is positive semi-definite.
The corresponding expected Laplacian matrix can be partitioned into the following form 
$$\mathbf{L}^{\rm M}(\mathcal{G})=\begin{bmatrix}
\mathbf{L}^{\rm M}_{aa} &\mathbf{L}^{\rm M}_{af} \\
\mathbf{L}^{\rm M}_{fa} &\mathbf{L}^{\rm M}_{ff} \\
\end{bmatrix},$$
where $\mathbf{L}_{aa}^{\rm M}$ is the block entry of matrix $\mathbf{L}^{\rm M}$ in the first $n_ad$ rows
and $n_ad$ columns.

\begin{lemma}
\text{\emph{\cite{bearingzhao}}}
 Suppose that $\mathcal{G}(\bar{\mathbf{x}})$ is infinitesimal bearing rigid, the matrix $\mathbf{L}^{\rm M}_{ff}$ is positive definite if and only if $n_a \geq 2$.
\end{lemma}

\begin{theorem}\label{NLmaintheorem}
  Suppose that $\mathcal{G}(\overline{\mathbf{x}})$ is infinitesimal bearing rigid and $n_a \geq 2$. Selecting the updating step size such that $\alpha<{\min}(\frac{1}{\max_{i\in V_f}\| \mathbf{D}^{\rm M}_{i} \|},\frac{2}{{\max_{i,j \in V}}||\mathbf{A}_{ij}||})$, the first and second moments of the system \eqref{algorithmNL13} converge to zero as $k\to \infty$, i.e., the estimated configuration ${\hat{\overline{\mathbf{x}}}}(k)$ converges to the actual configuration $\overline{\mathbf{x}}$ as $k \to \infty$ in probability. 
\end{theorem}

\emph{Proof:} See Appendix E.

{black}{\begin{remark}
Theorem \ref{NLmaintheorem} implies the rank condition $\text{rank}(\mathbf{L})=dn-d-1$. In this case, the expected convergence of the position estimate vector corresponds to a clustering consensus configuration \cite{THM2018} ($n$ clusters corresponding to $n$ agents). It is notable that this clustering phenomenon cannot be seen in the randomized scalar weighted consensus algorithms in \cite{boyd2006} and other subsequent works.  
\end{remark}}
{black}{\section{Application in formation control problem with position estimation}\label{section: FC}}

In this section, we extend our work to the problem of formation control for a group of agents without using knowledge of their real position in the inertial coordinate \cite{Oh2012TAC}. Based on only the information about the relative positions with neighbors, each agent is controlled to track its reference position, and the overall desired formation can be finally achieved, up to translation. The main advantage of this approach is that it does not require GPS devices, i.e., is cost-effective. Moreover, since the gossip-based communication protocol is applied, simplicity and reliability are guaranteed.

\subsection{Problem Formulation}\label{prob_FC}

Consider a group of $n$ agents (robots) in $d$ dimensional space where their dynamics can be modeled by a first-order integral system

\begin{equation}
    \overline{x}_i(k+1) = \overline{x}_i(k) + \overline{u}_i(k),
\end{equation}
where $\overline{x}_i(k), \overline{u}_i(k) \in \mathbb{R}^d$ denotes the absolute position and the control input of the $i^{th}$ agent at time instant $k$, respectively. Although the information of $\overline{x}_i(k)$ is not available, it is assumed that each agent can sense the relative positions of its neighbors
\begin{equation}
    \overline{x}_{ij}(k) = \overline{x}_j(k) - \overline{x}_i(k),
\end{equation}
where agent $j$ is a neighbor of agent $i$. Our system now can be modeled as a matrix weighted graph $ \mathcal{G}=(V,E,A)$, where, $V=\left\{ 1,2,...,n\right\}$ is the vertex set (agents/robots), $E\subseteq V\times V$ is the edge set, and 
$A=\{ \mathbf{A}_{ij} \in \mathbb{R}^{d \times d}|~(i,j) \in E\}$ denotes the set of matrix weights. 

The desired formation $\mathbf{\overline{x}}^*(k)$ is represented by the desired position for each agent $\overline{x}_i^*$. Denote $\mathbf{\overline{x}}(k) = \text{vec}
(\overline{x}_i(k))$, our problem is formulated as 

\textbf{Problem 2.} \emph{Suppose that the initial position of the system is $\hat{\overline{x}}(0)$. Design the control law for each agent $\overline{u}_i(k)$ based on the relative position information $\overline{x}_{ij}$ such that the desired formation can be achieved up to a translation, i.e., $\mathbf{\overline{x}}(k) \rightarrow \mathbf{\overline{x}}^* + \tilde{\mathbf{\overline{x}}}_{\infty}$ where $\tilde{\mathbf{\overline{x}}}_{\infty}$ is constant.}

\subsection{Randomized formation control with position estimation}

Consider the system in Subsection \ref{prob_FC} with the randomized distributed protocol installed. Each agent/robot contains a variable $\hat{\overline{x}}_i(k)$ representing the estimation of its actual position. At time slot $k$, suppose that agent $i$ wakes up, then it will choose only one neighbor $j$ with a probability ${\rm P}_{ij}$ to communicate. Then both agents $i$ and $j$ will update their estimation values and apply the proposed control law as follows
\begin{align}
    \hat{\overline{x}}_i(k+1) &= \hat{\overline{x}}_i(k) - \alpha \mathbf{A}_{ij}\big[\hat{\overline{x}}_i(k)-\hat{\overline{x}}_j(k)+\overline{x}_{ij}(k) \big] 
     + \overline{u}_i(k) \nonumber\\
    \hat{\overline{x}}_j(k+1) &= \hat{\overline{x}}_j(k)  - \alpha \mathbf{A}_{ij}\big[\hat{\overline{x}}_j(k)-\hat{\overline{x}}_i(k)+\overline{x}_{ji}(k)  \big] + \overline{u}_j(k) \nonumber\\
    \overline{u}_i(k) &=-\sigma(\hat{\overline{x}}_i(k)-\overline{x}_i^*) \nonumber\\
    \overline{u}_j(k) &=-\sigma(\hat{\overline{x}}_j(k)-\overline{x}_j^*), \label{FC_law}
\end{align}
where 
$\alpha, \sigma > 0$ are the common coefficients of the estimators and controllers that will be designed later. $\hat{\overline{x}}_i(k)$ is the position estimation vector of agent $i$ and $\overline{u}_i(k)$ is its feedback controller. Other $n-2$ agents of the group will not update their estimation values and will maintain their positions by applying zero control effort. 

System (\ref{FC_law}) can be rewritten as

\begin{equation}
    \begin{aligned}
     \mathbf{\hat{\overline{x}}}(k+1) &= \mathbf{\hat{\overline{x}}}(k) - {\alpha} \mathbf{L}_{ij}(\mathbf{\hat{\overline{x}}}(k)-\mathbf{\overline{x}}(k)) + \mathbf{\overline{u}}(k),\\
     \mathbf{\overline{x}}(k+1) &= \mathbf{\overline{x}}(k+1) + \mathbf{\overline{u}}(k),\\
     \mathbf{\overline{u}}(k) &= -{K_{ij}}(\mathbf{\hat{\overline{x}}}(k)-\mathbf{\overline{x}}^*(k)),
    \end{aligned}
\end{equation}
where $\mathbf{\hat{\overline{x}}}(k) = \text{vec}(\hat{\overline{x}}_i(k))$, $\mathbf{\hat{\overline{u}}}(k) = \text{vec}(\hat{\overline{u}}_i(k))$. The block matrix $\mathbf{L}_{ij}$ and the control gain {${K_{ij}}$} are defined as
{
\begin{equation}\label{Lij}
\begin{aligned}
\mathbf{L}_{ij}&=\begin{bmatrix}
    &\mathbf{0}       &\cdots     &\mathbf{0}            &\cdots    &\mathbf{0} &\cdots &\mathbf{0} \\
    &\vdots     &\ddots     &\vdots          &\ddots    &\vdots &\ddots &\vdots \\
    &\mathbf{0}       &\cdots  &\mathbf{A}_{ij} &\cdots    &-\mathbf{A}_{ij} &\cdots &\mathbf{0} \\
    &\vdots     &\ddots     &\vdots          &\ddots    &\vdots &\ddots &\vdots \\
    &\mathbf{0}      &\cdots  & -\mathbf{A}_{ij} &\cdots &\mathbf{A}_{ij} &\cdots &\mathbf{0}  \\
    &\vdots     &\ddots     &\vdots          &\ddots    &\vdots &\ddots &\vdots \\
    &\mathbf{0}       &\cdots     &\mathbf{0}            &\cdots    &\mathbf{0} &\cdots &\mathbf{0}
\end{bmatrix},\\
K_{ij} &= \text{blkdiag}(\mathbf{0},\dots, \sigma\mathbf{I}_d,\dots,\sigma\mathbf{I}_d,\dots,\mathbf{0}),
\end{aligned}
\end{equation}}

where $\sigma\mathbf{I}_d > 0$ are two block entries of matrix {${K_{ij}}$}, one lies in the $({(i-1)d+1:id})^{\text{th}}$ rows
and $({(i-1)d+1:id})^{\text{th}}$ columns and the other in the $({(j-1)d+1:jd})^{\text{th}}$ rows
and $({(j-1)d+1:jd})^{\text{th}}$ columns. Denote the position estimation error vector $\mathbf{\overline{e}}_1(k)=\mathbf{\hat{\overline{x}}}(k)-\mathbf{\overline{x}}(k)$ and the {position error} vector $\mathbf{\overline{e}}_2(k)=\mathbf{\overline{x}}(k)-\mathbf{\overline{x}}^*$. With a probability $\frac{1}{n}{\rm P}_{ij}$, we have
\begin{equation}\label{overall_FC}
\begin{aligned}
\mathbf{\overline{e}}_1(k+1) &= \mathbf{\hat{\overline{x}}}(k+1)-\mathbf{\overline{x}}(k+1)\\
&= (\mathbf{I}_{dn}-{\alpha}\mathbf{L}_{ij})\mathbf{\overline{e}}_1(k) 
= W_{ij} \mathbf{\overline{e}}_1(k),\\
\mathbf{\overline{e}}_2(k+1) &= 
\mathbf{\overline{x}}(k)-{{K_{ij}}}(\mathbf{\hat{\overline{x}}}(k)-\mathbf{\overline{x}}^*(k))-\mathbf{\overline{x}}^*\\
&= (\mathbf{I}_{dn}-{{K_{ij}}})\mathbf{\overline{e}}_2(k) - {{K_{ij}}}\mathbf{\overline{e}}_1(k) ,
\end{aligned}
\end{equation}
where the expression of $W_{ij}$ is the same as that in (\ref{algorithm11}). We now utilize the definition of the expected matrix weighted graph, taking the expectation of (\ref{overall_FC}) lead to
\begin{equation}\label{FC_sys}
    \begin{aligned}
        {\rm E}[\mathbf{\overline{e}}_1(k+1)] &= (\mathbf{I}_{dn}-\alpha \mathbf{L}^{\rm M}){\rm E}[\mathbf{\overline{e}}_1(k)]\\
        {\rm E}[\mathbf{\overline{e}}_2(k+1)]&= (\mathbf{I}_{dn}-\overline{K}){\rm E}[\mathbf{{\overline{e}}}_2(k)]-\overline{K}{\rm E}[\mathbf{\overline{e}}_1(k)],
    \end{aligned}
\end{equation}
where $\overline{K}=
\sum_{i,j=1}^n\frac{1}{n}{\rm P}_{ij}K_{ij}$ 
is a block diagonal matrix whose entries {$\frac{\sigma}{n}\sum_{j}({\rm P}_{ij}+{\rm P}_{ji})$} in its $({(i-1)d+1:id})^{\text{th}}$ rows
and $({(i-1)d+1:id})^{\text{th}}$ columns. We have the following theorem
\begin{theorem}\label{FC_convergence}
    Suppose that there exists a positive spanning tree in $\mathcal{G}$. If the control and estimator gains are chosen such that $\alpha < {\min}(\frac{1}{\max_{i\in V}\| \mathbf{D}^{\rm M}_{i} \|},\frac{1}{\max_{i,j\in V} \| \mathbf{A}_{ij} \|}), {\sigma} < \frac{2n}{\sum_{j}({\rm P}_{ij}+{\rm P}_{ji})}$, the actual position of agents will eventually converge in expectation to the desired formation up to an unknown constant  translation $\mathbf{\Tilde{\overline{x}}}_{\infty}=-\frac{1}{n}(\mathbf{1}_n\mathbf{1}_n^\top {\otimes \mathbf{I}_d })\mathbf{\overline{e}}_1(0)$.
\end{theorem}
\emph{Proof:} See Appendix F.
\begin{remark}
    Although the actual positions of agents are assumed to be unknown, if the information about the initial center of these agents $\frac{1}{n}(\mathbf{1}_n\mathbf{1}_n^\top {\otimes \mathbf{I}_d})\mathbf{\overline{x}}(0)$ is available, one can initialize the values of the estimator such that $\frac{1}{n}(\mathbf{1}_n\mathbf{1}_n^\top {\otimes \mathbf{I}_d})\mathbf{\hat{\overline{x}}}(0)=\frac{1}{n}(\mathbf{1}_n\mathbf{1}_n^\top {\otimes \mathbf{I}_d})\mathbf{\overline{x}}(0)$, i.e., $\mathbf{\overline{e}}_1(0)= \mathbf{0}_{dn \times 1}$. 
    As a result, the desired formation can be achieved in expectation with zero tracking error.
\end{remark}
\section{Numerical examples}
\label{section: simulation}
\subsection{Randomized matrix weighted consensus  with Leaderless topology} \label{sim:leaderless}
\begin{figure}
\begin{center}
\includegraphics[height=6cm]{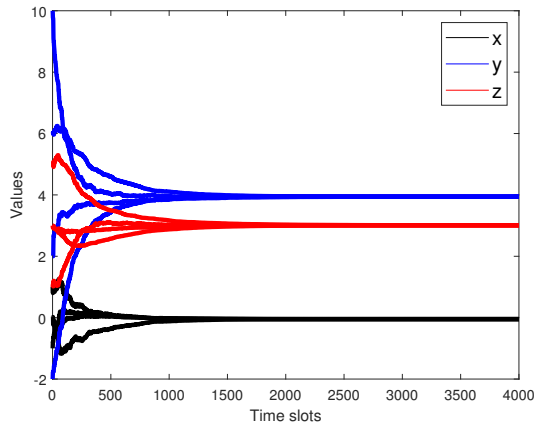}    
\caption{Randomized Randomized matrix weighted of 4 agents with
Leaderless topology.}  
\label{leaderless}                                 
\end{center}                                 
\end{figure}

Consider a system of four agents whose state vectors are defined in $\mathbb{R}^3$. The matrix weights of the system are given as
{\small
\begin{align*}
\mathbf{A}_{12}&= 
\begin{bmatrix}
1 &0 &0\\
0 &\frac{1}{2} &\frac{1}{5}\\
0 &\frac{1}{2} &1
\end{bmatrix},
\mathbf{A}_{21}= 
\begin{bmatrix}
4 &2 &0\\
2 &\frac{1}{2} &0\\
0 &0 &0
\end{bmatrix}, 
\mathbf{A}_{13}= 
\begin{bmatrix}
5 &\frac{1}{3} &0\\
\frac{1}{3} &\frac{1}{2} &0\\
0 &0 &0
\end{bmatrix},\\
\mathbf{A}_{31}&= 
\begin{bmatrix}
5 &\frac{1}{3} &0\\
\frac{1}{3} &\frac{1}{2} &0\\
0 &0 &1
\end{bmatrix},
\mathbf{A}_{23} = 
\begin{bmatrix}
1 &\frac{1}{2} &0\\
\frac{1}{2} &1 &0\\
0 &0 &\frac{1}{3}
\end{bmatrix},
\mathbf{A}_{32}= 
\begin{bmatrix}
1 &\frac{1}{2} &0\\
\frac{1}{2} &\frac{1}{2} &0\\
0 &0 &\frac{1}{3}
\end{bmatrix},\\
\mathbf{A}_{34} &= 
\begin{bmatrix}
2 &0 &\frac{1}{2}\\
0 &2 &0\\
\frac{1}{2} &0 &1
\end{bmatrix},
\mathbf{A}_{43}= 
\begin{bmatrix}
3 &0 &0\\
0 &1 &1\\
0 &1 &1
\end{bmatrix}.
\end{align*}
}
and zero matrices otherwise. The transition matrix $\pi = [{\rm P}_{ij}]$ is designed as follow
{
\begin{equation*}
\pi =  
\begin{bmatrix}
0 &\frac{1}{2} &\frac{1}{2} &0\\
\frac{1}{3} &0 &\frac{2}{3} &0\\
\frac{1}{2} &\frac{1}{4} &0 &\frac{1}{4}\\
0 &0 &1 &0
\end{bmatrix}.
\end{equation*}}
The initial vectors of the agents are given as $\overline{x}_1(0)=[1,-2,3]^\top, \overline{x}_2(0)=[0,6,3]^\top,\overline{x}_3(0)=[-1,10,1]^\top,\overline{x}_3(0)=[0,2,5]^\top$. The average state vector is thus $\hat{\overline{x}}=[0,4,3]^\top$. To meet the requirements of Theorem~\ref{upperbound1}, the agents' common step size is set at $\alpha=0.02$. As shown in Fig.~\ref{leaderless}, four agents achieve a consensus. Their convergence value is $[0.0,4.0,3.0]^\top$, which is the average.



\subsection{Randomized matrix weighted consensus with Leader-Following topology}
\begin{figure}
\begin{center}
\includegraphics[height=6cm]{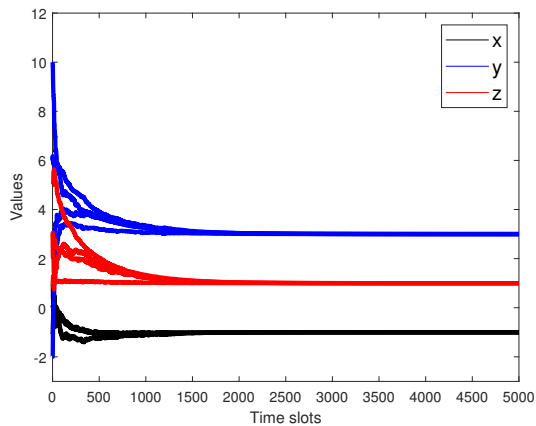}    
\caption{Randomized Randomized matrix weighted of 4 agents with
Leader-Following topology}  
\label{leader}                                 
\end{center}                                 
\end{figure}

We added a leader node $\overline{x}_0=[-1,3,1]^\top$ to the previous system. The leader-follower matrix-weights are given as { $ \mathbf{A}_{10}=
\begin{bmatrix}
1 & 0 & 0\\
0 & 1 & 0\\
0 & 0 & 0
\end{bmatrix},
\mathbf{A}_{20}=
\begin{bmatrix}
0 & 0 & 0\\
0 & 0 & 0\\
0 & 0 & 5
\end{bmatrix}
$} 
and zero matrices, otherwise. The common step size is set at $\alpha = 0.1, \theta=0.5$. The simulation result is depicted in Fig.~\ref{leader}. All agents asymptotically reach a consensus at the leader's state vector $[-1,~3,~1]^\top$.

\subsection{Randomized bearing-based network localization}\label{NL}

\begin{figure*}[ht]
    \centering
    \begin{minipage}{0.5\linewidth} 
    \subfloat[Network $\mathcal{G}(\mathbf{x})$]{\includegraphics[width=0.45\textwidth]{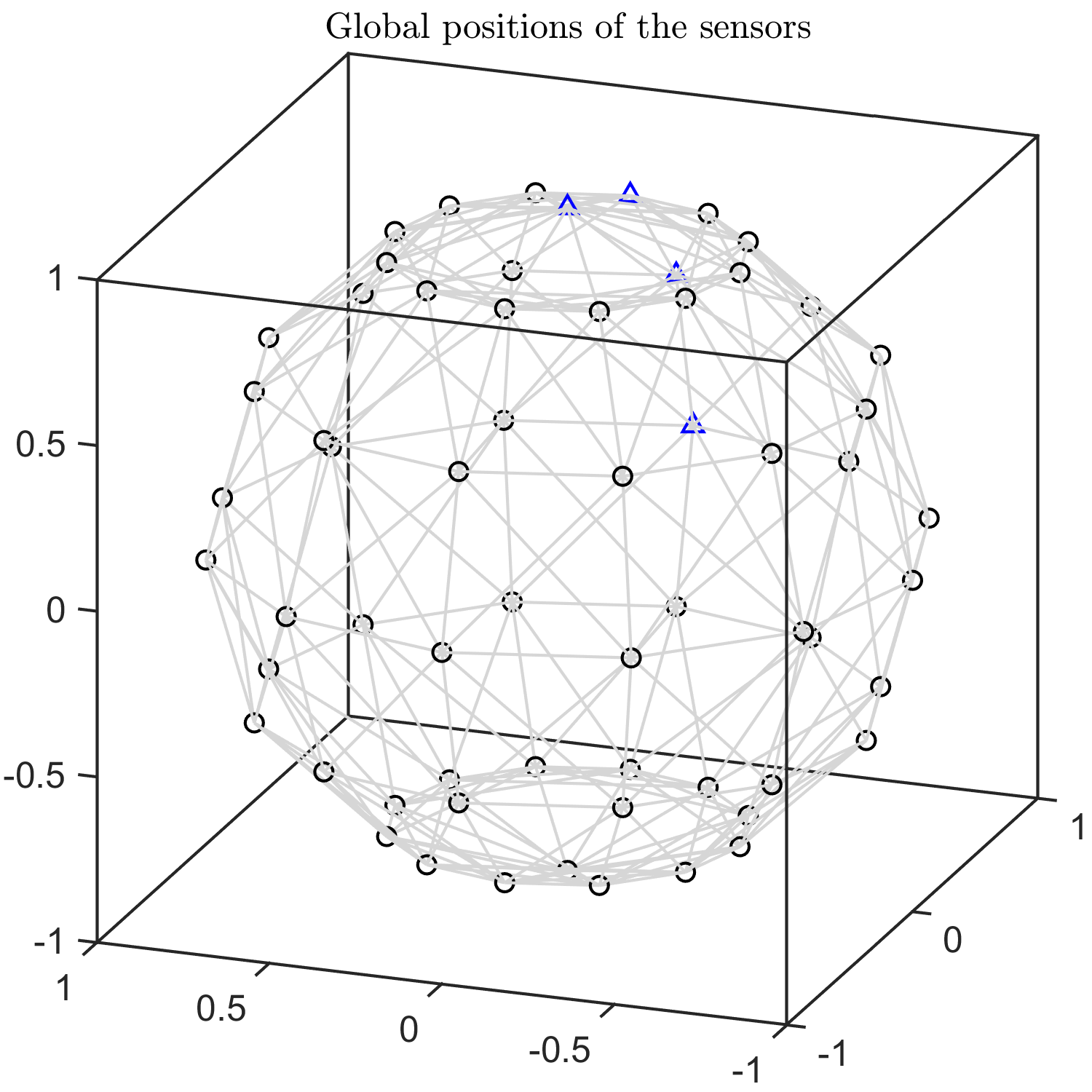} \label{trueConfig}} \hfill
    \subfloat[$\hat{\overline{x}}(0)$]{\includegraphics[width=0.45\textwidth]{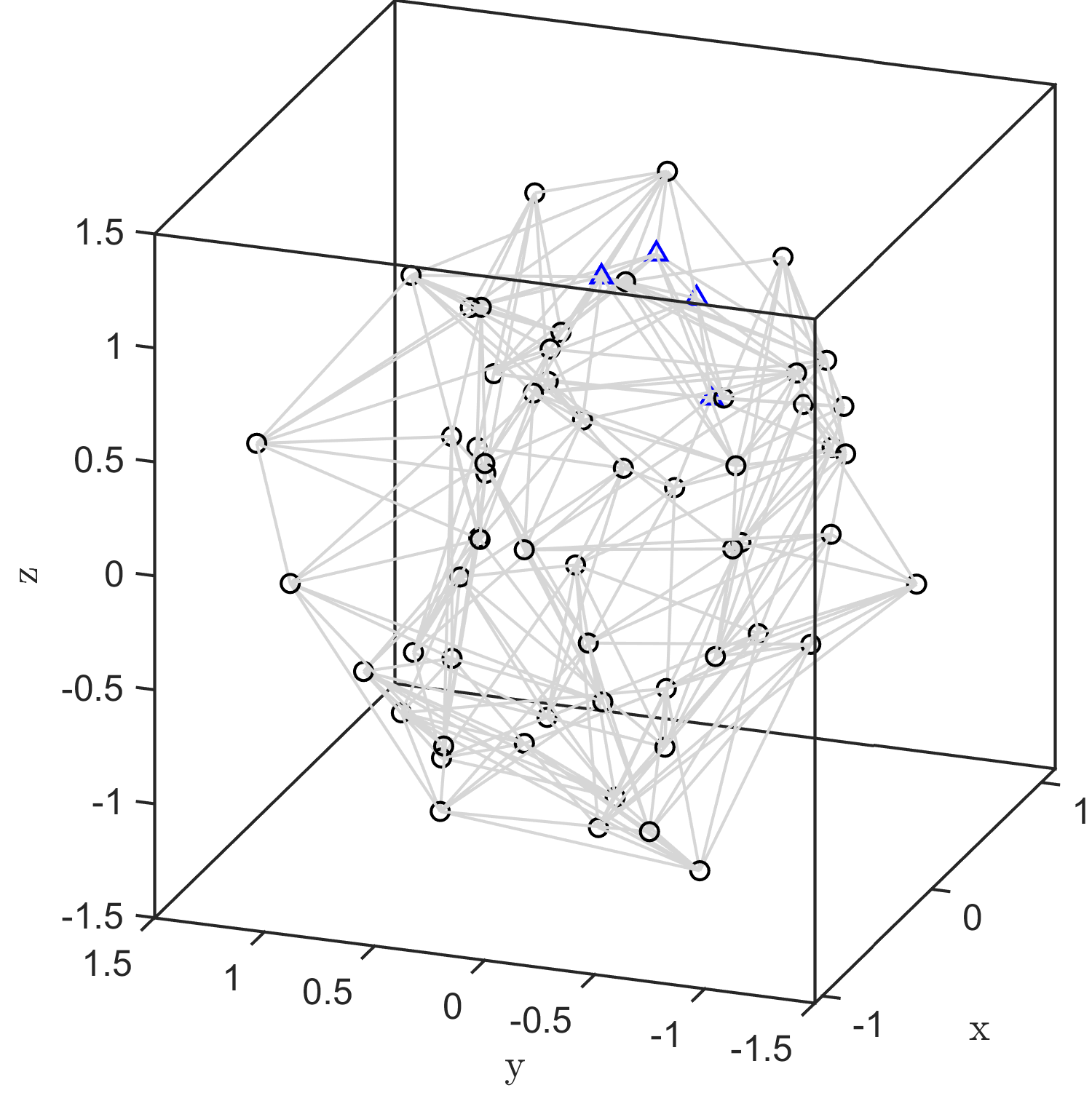}\label{est0}}\hfill
    \subfloat[$\hat{\overline{x}}(100)$]{\includegraphics[width=0.45\textwidth]{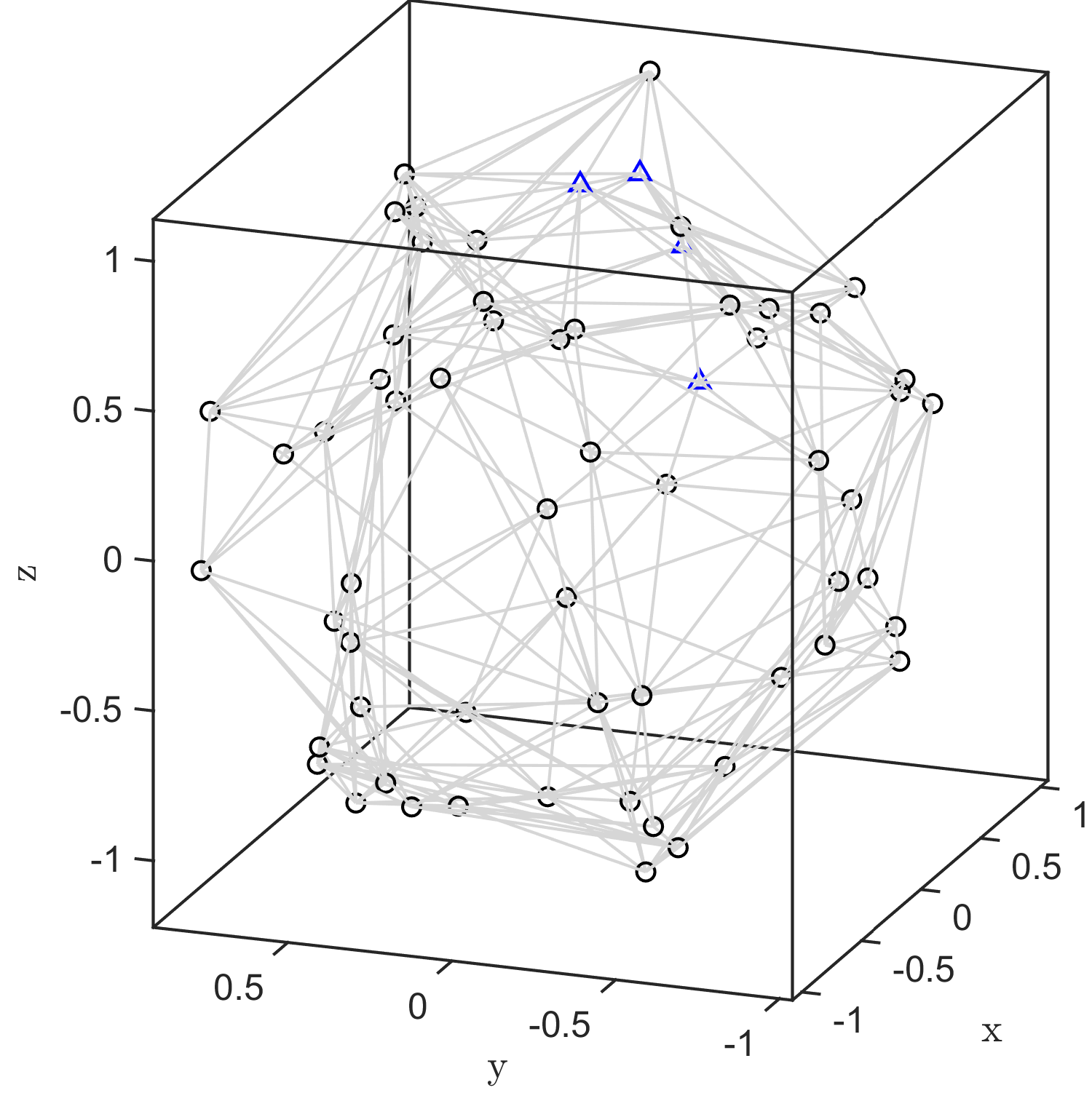}\label{est100}} \hfill
    \subfloat[$\hat{\overline{x}}(300)$]{\includegraphics[width=0.45\textwidth]{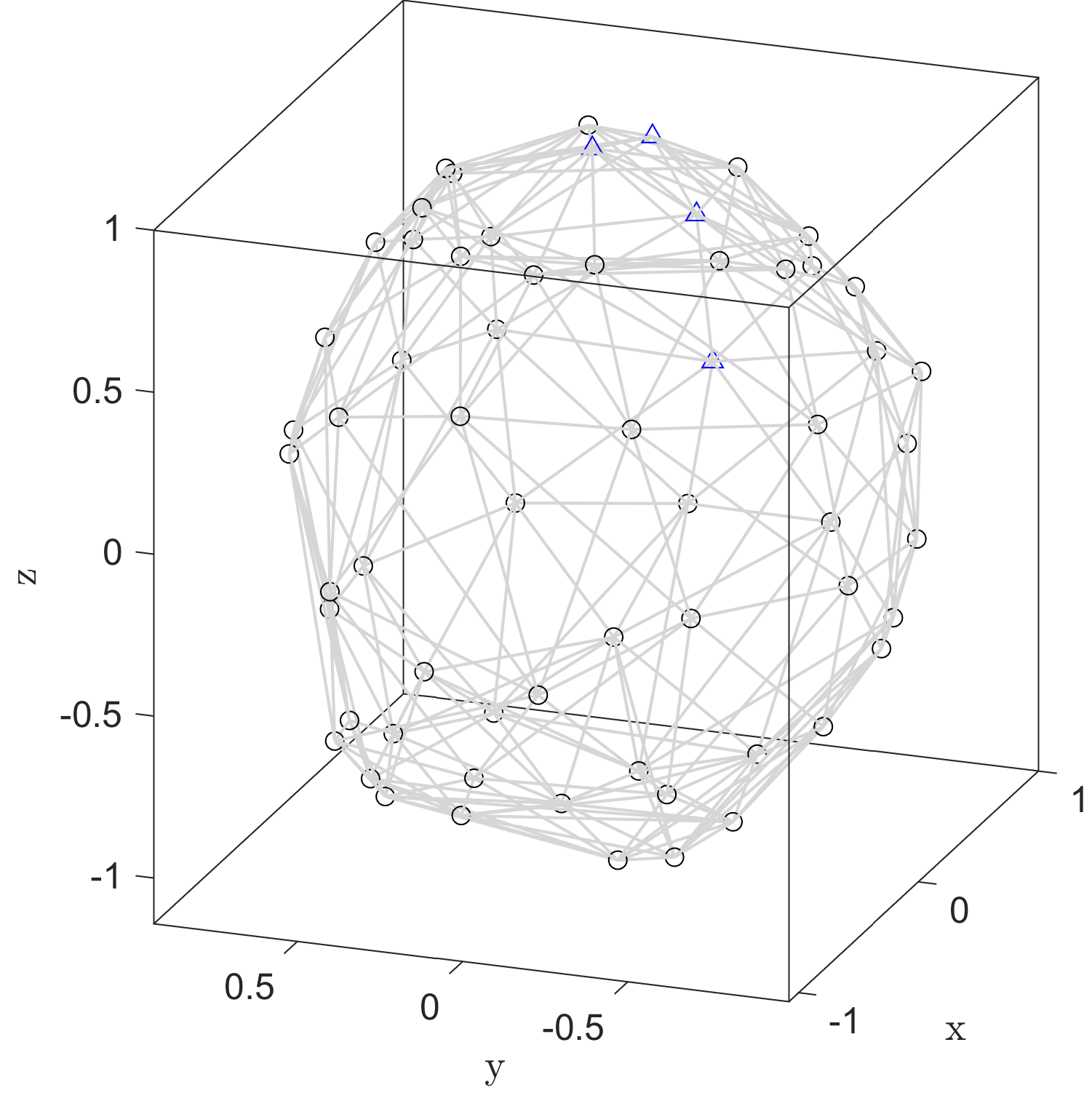}\label{est300}} \hfill
    \subfloat[$\hat{\overline{x}}(500)$]{\includegraphics[width=0.45\textwidth]{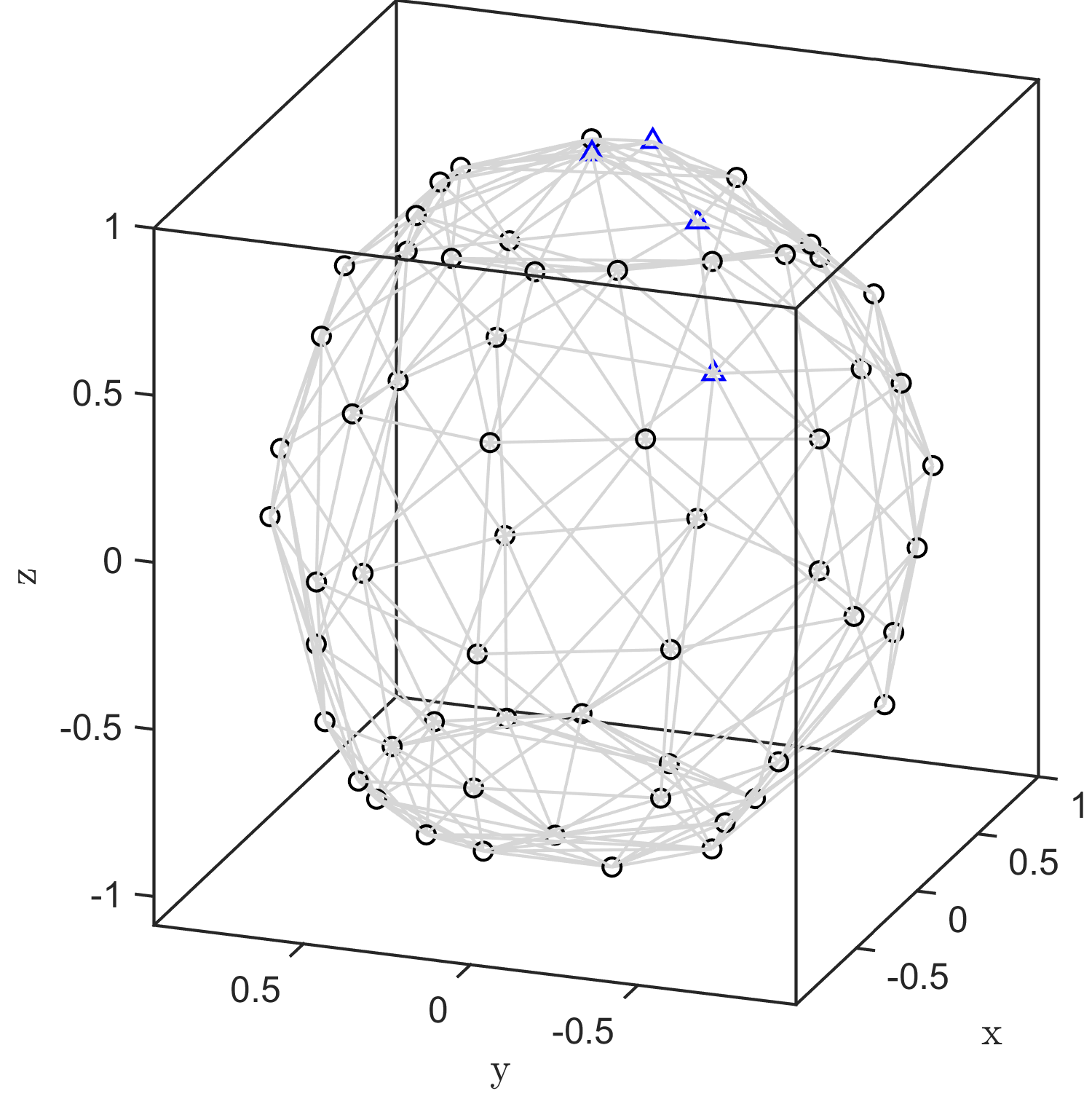}\label{est500}}\hfill
    \subfloat[$\hat{\overline{x}}(1000)$]{\includegraphics[width=0.45\textwidth]{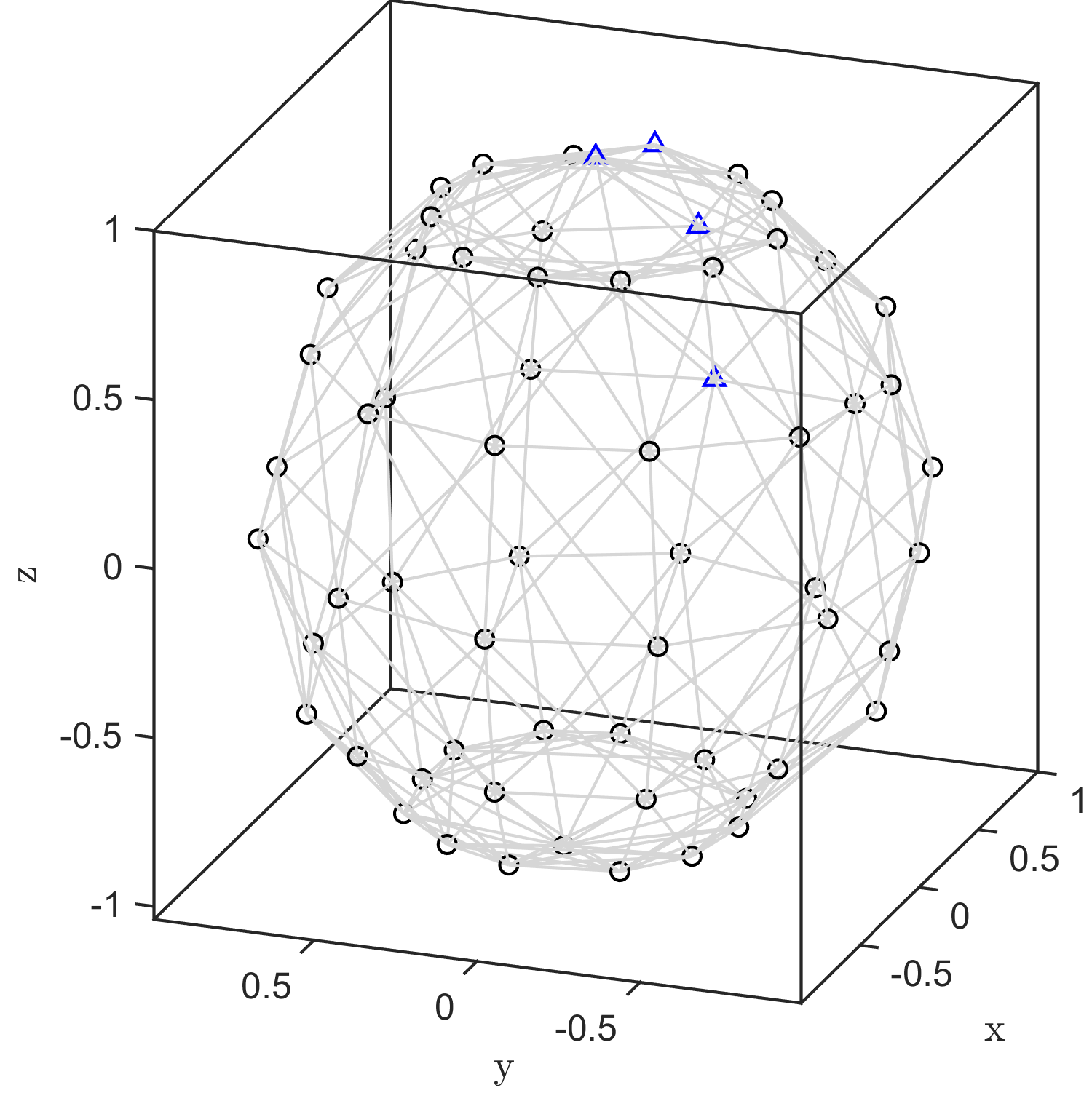}\label{est1000}}
    \end{minipage}%
    \hfill
    \begin{minipage}{0.4\linewidth}
    \centering
    \subfloat[Bearing error vs time.]{\includegraphics[width=1\textwidth]{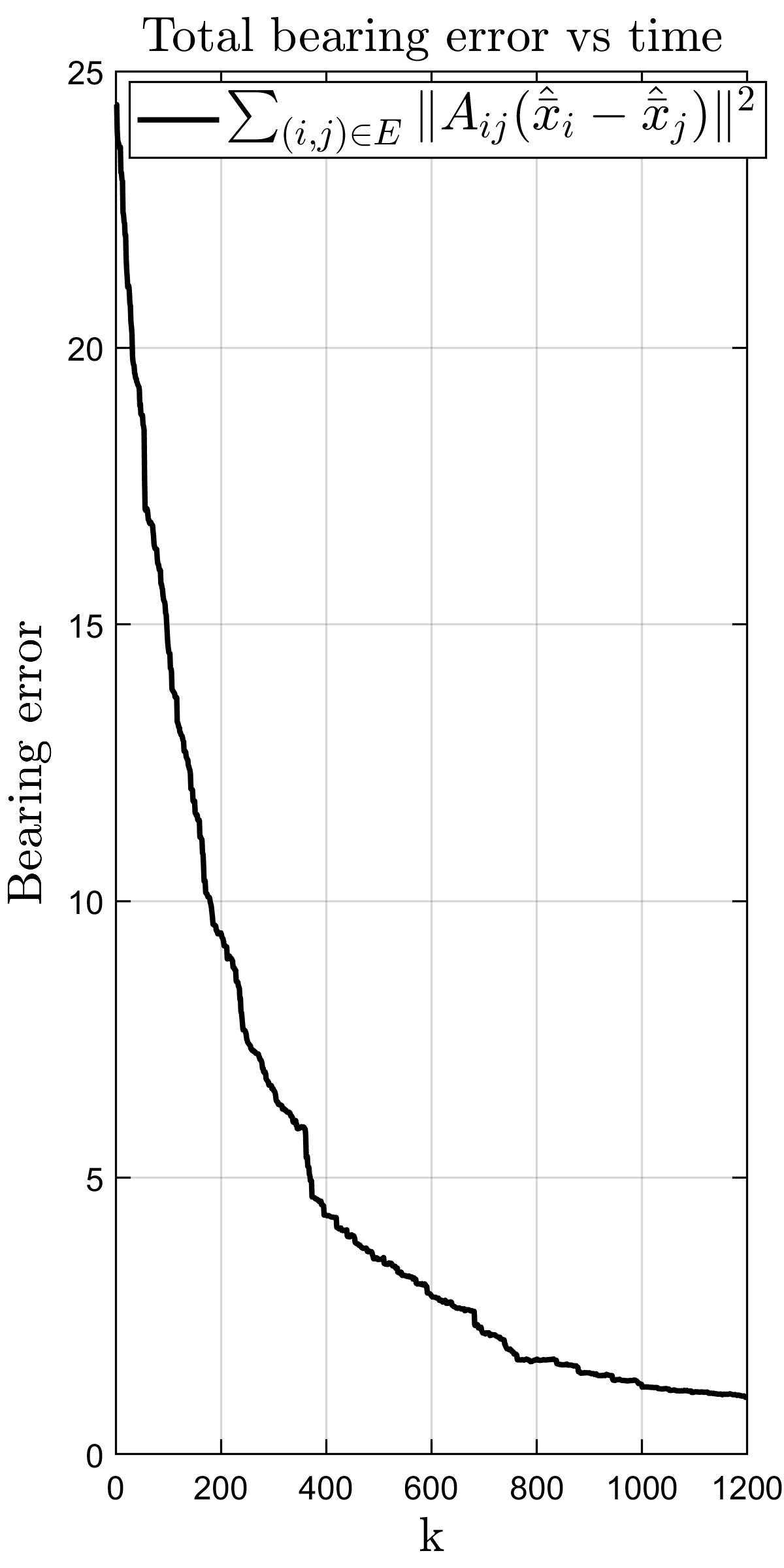}\label{Berror}}
    \end{minipage}%
    \caption{Simulation of a sensor network consisting of 62 nodes under the gossip-based network localization protocol \eqref{algorithmNL0}, \eqref{algorithmNL1}, \eqref{algorithmNL2}: (a) - the network $\mathcal{G}(\bar{\mathbf{x})}$ (beacon nodes are denoted with `${black}{\boldsymbol{\Delta}}$', normal nodes are denoted by `\textbf{o}', respectively); From (b) to (f) - the estimate configurations at different time instances; (g) the bearing error vs time. \label{fig:simNL}} 
\end{figure*}

Consider a network of $n=62$ sensor nodes in a three-dimensional space ($d=3$), with $\overline{x}_i=[x_i,y_i,z_i]^{\top}$. There are $n_a=4$ beacons (nodes 1 to 4) in the network. As depicted in Fig.~\ref{trueConfig}, the sensor nodes are distributed around a sphere of radius 1.

The initial estimate $\hat{\overline{x}}_i(0)=[\hat{x}_i(0),\hat{y}_i(0),\hat{z}_i(0)]^\top$ of each follower node is generated randomly in a cubic $[-1.5,1.5]\times[-1.5,1.5]\times[-1.5,1.5]$, which is shown in Fig. \ref{est0}. The edges in $E$, being chosen accordingly to the proximity-rule
\[((i,j) \in E) \longleftrightarrow (\|\bar{x}_i - \bar{x}_j\|\leq 0.75),\]
results to the topological graph $\mathcal{G}$ in Fig.~\ref{trueConfig}.

The simulation result of the sensor network under the randomized network localization protocol \eqref{algorithmNL0}, \eqref{algorithmNL1}, \eqref{algorithmNL2} is illustrated in Fig.~\ref{fig:simNL}. As can be shown in Fig.~\ref{est0}--\ref{est1000}, the estimate configuration $\hat{\overline{x}}(k)$ at time instances $k=0, 100, 300, 500, 1000$, demonstrates that all position estimates eventually converge to the true value as $k$ increases. 

Additionally, it can be seen from Fig.~\ref{Berror} that the total bearing error, which is defined as $\sum_{(i,j)\in E} \|\mathbf{A}_{ij}(\hat{\overline{x}}_j(k)-\hat{\overline{x}}_i(k))\|^2$, converges to 0. Thus, the simulation result is consistent with the theoretical results. 

\subsection{Randomized formation control with position estimation}
\begin{figure}
\begin{center}
\includegraphics[height=6cm]{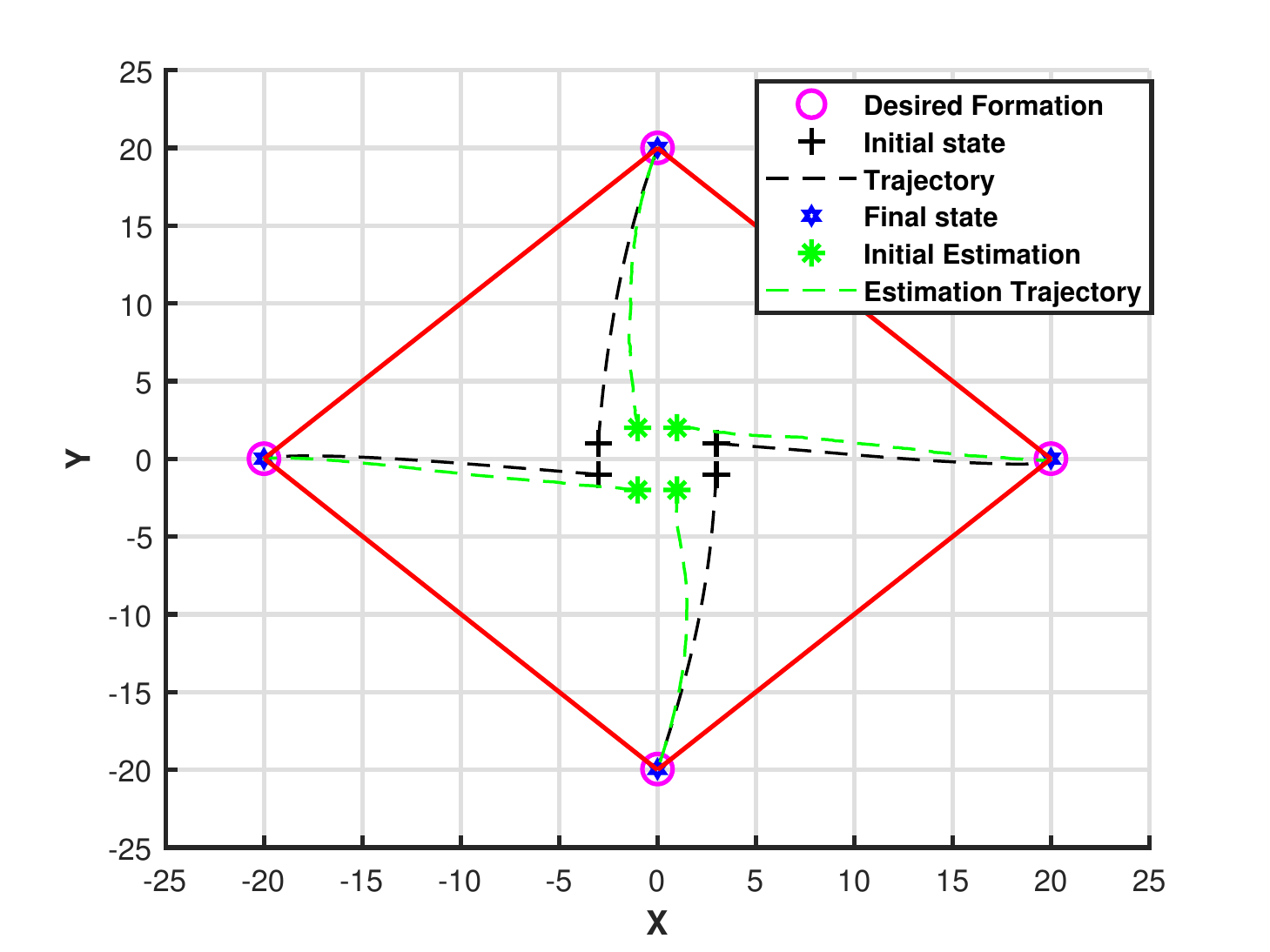}    
\caption{Trajectories of all agents under proposed control law}  
\label{FC}                                 
\end{center}                                 
\end{figure}

Consider a group of 4 agents/robots who are working in planar space. The initial positions of agents are given as $\overline{x}_1(0)=[-3,-1,0]^\top$, $\overline{x}_2(0)=[-3,1,0]^\top$, $\overline{x}_3(0)=[3,1,0]^\top$, $\overline{x}_4(0)=[3,-1,0]^\top$, and the desired formation is given as $\overline{x}_1^*=[-20,0,0]^\top$, $\overline{x}_2^*=[0,20,0]^\top$, $\overline{x}_3^*=[20,0,0]^\top$, $\overline{x}_4^*=[-20,0,0]^\top$. Initially, the position estimation values are given as $\hat{\overline{x}}_1(0)=[-1,-2,0]^\top, \hat{\overline{x}}_2(0)=[-1,2,0]^\top$, $\hat{\overline{x}}_3(0)=[3,1,0]^\top,\hat{\overline{x}}_4(0)=[3,-1,0]^\top$. The matrix weights among agents and the probability distribution are chosen exactly the same as those in Subsection \ref{sim:leaderless}. Fig.~\ref{FC} illustrates the trajectories of all agents and the estimation of their positions with $\alpha = 0.02, k = 0.01$. Overall, the estimator can eventually converge to the agents's actual positions. Additionally, the desired formation is tracked successfully.

\section{Conclusion}\label{sec:conclusion}
In this paper, the randomized matrix-weighted consensus algorithm are proposed for both leaderless and leader-following topologies. {Under some mild assumptions on the matrix weights, we showed that by choosing a small enough updating step-size, the agents will achieve consensus in expectations.} Furthermore, the upper bound of the convergence rate could be predetermined given the knowledge of each agent's probability distribution. The proposed framework is then applied to bearing-based network localization and formation control problems. Finally, numerical examples verified the theoretical result. One of our future research directions is to apply optimization techniques to find the optimal {step-size} or probability distribution for each agent, which will improve the consensus speed of the system. {Another promising direction for future studies is to extend the work in this paper to high-order multi-agent systems.}


%
\appendices

\section{Proof of Lemma \ref{stepsizeav}}
1) \emph{The set of all real symmetric matrices with eigenvalues in $[0,1]$ is convex. Every possible matrix $W_{ij}^\top W_{ij}$ is in this set.}

The first part of this statement is a consequence of the following lemma \cite{Rajendramatrix2}: {Let} \emph{A and B be n×n Hermitian matrices. If the eigenvalues of A and B are all in an interval I, then the eigenvalues of any convex combination of A and B are also in I.} 

Rearrange the rows and columns of $W_{ij}$ we get
\begin{equation}
W_{ij}=\begin{bmatrix}
        &\mathbf{I}_d-\alpha_i \mathbf{A}_{ij} &\alpha_i \mathbf{A}_{ij} &\cdots &\mathbf{0}  \\
        &\alpha_j \mathbf{A}_{ij} &\mathbf{I}_d-\alpha_j\mathbf{A}_{ij} &\cdots &\mathbf{0}  \\
       &\vdots          &\vdots    &\ddots  &\vdots \\
        &\mathbf{0}   &\mathbf{0}    &\cdots &\mathbf{I}_d  \\
\end{bmatrix}.
\end{equation}
Thus, the union of the unity eigenvalues (which correspond to the identity matrices $\mathbf{I}_d$) and the eigenvalues of matrix $w=
\begin{bmatrix}
    &\mathbf{I}_d-\alpha_i \mathbf{A}_{ij} &\alpha_i \mathbf{A}_{ij}\\
    &\alpha_j \mathbf{A}_{ij} &\mathbf{I}_d-\alpha_j\mathbf{A}_{ij} \\
\end{bmatrix}
$ yields the eigenvalues of $W_{ij}$. By choosing $\alpha<\frac{1}{\max_{i,j\in V}\lVert \mathbf{A}_{ij} \rVert}$, we can easily obtain $-1<\lambda(w)\leq 1$ (see Lemma \ref{stepsize}). $W_{ij}$'s eigenvalues then satisfy $\lambda(W_{ij}) \in (-1,1]$. Hence, $\lambda(W_{ij}^{\top}W_{ij})=\lambda(W_{ij}^2)=\lambda(W_{ij})^2 ~\in [0,1]$.

{\flushleft2) \emph{${\rm E}[W(k)^\top W(k)]$ has a unity spectral radius, and its unity eigenvalues are semi-simple.}}

We take into account ${\rm E}[W(k)^{\top}W(k)]$  as a convex combination of all possible $W_{ij}^2$. As a result, using the first statement of Lemma \ref{stepsizeav}, ${\rm E}[W(k)^\top W(k)]$ has a unity spectral radius. Due to the symmetry of ${\rm E}[W(k)^{\top}W(k)]$,
its unity eigenvalues are semi-simple.

{\flushleft3) \emph{$\mathbf{1}_n \otimes\mathbf{I}_d$ are 
only $d$ orthogonal right eigenvectors corresponding to the unity eigenvalues $\lambda=1$ of ${\rm E}[W(k)^\top W(k)]$ if and only if null$(\mathbf{L}^{\rm M})=\text{span}(\mathbf{1}_n\otimes \mathbf{I}_d)$}.}

Using (\ref{algorithm11}), we can easily get the following:
{
\begin{equation*}
W_{ij}^{\top}W_{ij}=\begin{bmatrix}
    &\mathbf{I}_d        &\cdots     &\mathbf{0}             &\cdots    &\mathbf{0} &\cdots &\mathbf{0} \\
    &\vdots     &\ddots     &\vdots          &\ddots    &\vdots &\ddots &\vdots \\
    &\mathbf{0}        &\cdots  &\mathbf{I}_d-\mathbf{\hat{A}}_{ij} &\cdots    &\mathbf{\hat{A}}_{ij} &\cdots &\mathbf{0}  \\
    &\vdots     &\ddots     &\vdots          &\ddots    &\vdots &\ddots &\vdots \\
    &\mathbf{0}        &\cdots  &\mathbf{\hat{A}}_{ij} &\cdots    &\mathbf{I}_d-\mathbf{\hat{A}}_{ij} &\cdots &\mathbf{0}  \\
    &\vdots     &\ddots     &\vdots          &\ddots    &\vdots &\ddots &\vdots \\
    &\mathbf{0}        &\cdots     &\mathbf{0}             &\cdots    &\mathbf{0} &\cdots &\mathbf{I}_d 
\end{bmatrix}
\end{equation*}}
where $\mathbf{\hat{A}}_{ij}=2\alpha \mathbf{A}_{ij}(\mathbf{I}_d-\alpha \mathbf{A}_{ij})$. Denote $\mathbf{\hat{M}}_{ij}= \frac{1}{n}(\mathbf{\hat{A}}_{ij} {\rm P}_{ij}+\mathbf{\hat{A}}_{ji} {\rm P}_{ji})$. By choosing $0<\alpha<\frac{1}{\max_{i,j\in V}\lVert \mathbf{A}_{ij} \rVert}$, $\mathbf{\hat{A}}_{ij}$ is positive definite (resp., positive semi-definite) if and only if $\mathbf{A}_{ij}$ is positive definite (resp., positive semi-definite). The same kind of relationship also holds for $\mathbf{\hat{M}}_{ij}$ and $\mathbf{M}_{ij}$. Moreover, it is simply provable that $\text{null}(\mathbf{\hat{A}}_{ij})=\text{null}(\mathbf{A}_{ij})$ and thus $\text{null}(\mathbf{\hat{M}}_{ij})=\text{null}(\mathbf{M}_{ij})$. We now rewrite ${\rm E}[W(k)^\top W(k)]$ as 
$${\rm E}[W(k)^\top W(k)]=\mathbf{I}_{dn} - (\mathbf{\hat{D}}^{\text{M}}-\mathbf{\hat{A}}^\text{M})=\mathbf{I}_{dn} - \mathbf{\hat{L}}^\text{M},$$
in which $\mathbf{\hat{D}}^{\text{M}}= \text{blkdiag}(\mathbf{\hat{D}}^{\text{M}}_{1},\ldots,\mathbf{\hat{D}}^{\text{M}}_{n})$, whereas $\mathbf{\hat{D}}^{\text{M}}_{i}= \sum_{j \in V}\mathbf{\hat{M}}_{ij}$ and $\mathbf{\hat{A}}^\text{M}=[\mathbf{\hat{M}}_{ij}]$. It is clear that the eigenvectors corresponding to the unity eigenvalue of ${\rm E}[W(k)^\top W(k)]$ {spans} the nullspace of $\mathbf{\hat{L}}^\text{M}$. Similar to Lemma \ref{nullspace}, the following holds:
  $\text{null}(\mathbf{\hat{L}}^{\text{M}}) =  \text{span}\{ \text{range}(\mathbf{1}_n \otimes \mathbf{I}_d),  \{ \mathbf{v}=[v_1^\top,\dots, v_n^\top]^\top \in \mathbb{R}^{nd}|~(v_j-v_j)\in \text{null}(\mathbf{\hat{M}}_{ij})=\text{null}(\mathbf{M}_{ij}), \forall (i,j) \in V \} \}$. Therefore, $\text{null}(\mathbf{\hat{L}}^{\text{M}})=\text{range}(\mathbf{1}_n\otimes \mathbf{I}_d)$ if and only if $\text{null}(\mathbf{L}^{\text{M}})=\text{range}(\mathbf{1}_n\otimes \mathbf{I}_d)$.

\section{Proof of Lemma~\ref{iff}}
The proof is divided into two parts.

{\flushleft1) \emph{Convergence of first moment.}} Take the expectation of both sides of equation (\ref{y}), the following yields
\begin{equation}\label{lemma51}
\begin{aligned}
    {\rm E}[\mathbf{\overline{y}}(k+1)]=\overline{W} {\rm E}[\mathbf{\overline{y}}(k)]=\overline{W}^{k+1}\mathbf{\overline{y}}(0).
\end{aligned}
\end{equation}
{Since $(\mathbf{1}_{n}^{\top}\otimes \mathbf{I}_n){\rm E}[\mathbf{\overline{y}}(k)]={\rm E}[(\mathbf{1}_{d}^{\top}\otimes \mathbf{I}_d)\mathbf{\overline{y}}(k)] = \mathbf{0}$, the convergence of first moment implies the average consensus of $\overline{y}_i(k)$. Hence, (\ref{lemma51}) 
has the form of a fast linear iteration system for distributed averaging \cite{boyd2006,XIAO2004}.} The convergence in the first moment, i.e., ${\rm E}[\mathbf{\overline{y}}(k+1)] \xrightarrow{k\rightarrow \infty} \frac{1}{n}\big((\mathbf{1}_n\mathbf{1}_n^\top)\otimes \mathbf{I}_d\big)\mathbf{\overline{y}}(0)$ is  equivalent to \cite{XIAO2004}

\begin{equation}
\begin{aligned}
\overline{W}^k\xrightarrow{k \rightarrow \infty} \frac{1}{n}\big((\mathbf{1}_n\mathbf{1}_n^\top)\otimes \mathbf{I}_d\big),
\end{aligned}
\end{equation}
{which can be rewritten as
\begin{equation}\label{Lemma53}
\begin{aligned}
\overline{W}^k- \frac{1}{n}\big((\mathbf{1}_n\mathbf{1}_n^\top)\otimes \mathbf{I}_d\big) \rightarrow \mathbf{0}_{dn\times dn},
\end{aligned}
\end{equation}
as $k \rightarrow \infty$. Considering the left-hand side of (\ref{Lemma53}), we have}

\begin{align}
\overline{W}^k-\frac{1}{n}\big((\mathbf{1}_n\mathbf{1}_n^\top)\otimes \mathbf{I}_d\big) &= \overline{W}^k ( \mathbf{I}_{dn} - \frac{1}{n}\big(\mathbf{1}_n\mathbf{1}_n^\top)\otimes \mathbf{I}_d\big ) \nonumber\\  
&= \overline{W}^k ( \mathbf{I}_{dn} - \frac{1}{n}\big(\mathbf{1}_n\mathbf{1}_n^\top)\otimes \mathbf{I}_d\big )^k \nonumber\\  
&= ( \overline{W} \big( \mathbf{I}_{dn} - \frac{1}{n}\big(\mathbf{1}_n\mathbf{1}_n^\top)\otimes \mathbf{I}_d\big) \big )^k \nonumber\\ 
&= ( \overline{W}  - \frac{1}{n}\big(\mathbf{1}_n\mathbf{1}_n^\top)\otimes \mathbf{I}_d\big) ^k ,\label{lemma 54}
\end{align}
where the second equality of (\ref{lemma 54}) can be proven easily using the induction method. {Meanwhile, the first and last inequalities can be obtained from the fact that $\overline{W}(\mathbf{1}_{n}\otimes \mathbf{I}_d)=\mathbf{1}_{n}\otimes \mathbf{I}_d$ (see (\ref{wbar}) and Lemma \ref{nullspace}).}  Now it is clear that the convergence of the first moment can be achieved if and only if matrix $( \overline{W}  - \frac{1}{n}\big(\mathbf{1}_n\mathbf{1}_n^\top)\otimes \mathbf{I}_d\big)$ has all eigenvalues lie inside the unit circle, i.e., Lemma \ref{iff}.

{\flushleft2) \emph{Convergence of the second moment}.} It is worth noting that as our main results are finding sufficient conditions for the convergence of the proposed consensus algorithm, the dynamic of ${\rm E}[\mathbf{\overline{y}}(k+1)^{\top}\mathbf{\overline{y}}(k+1)]$ matters \cite{boyd2006}. In order to obtain the necessary
and sufficient condition, considering the evolution of ${\rm E}[\mathbf{\overline{y}}(k+1)\mathbf{\overline{y}}(k+1)^{\top}]$ rather than ${\rm E}[\mathbf{\overline{y}}(k+1)^{\top}\mathbf{\overline{y}}(k+1)]$. From \cite{boyd2006}, we have
{\begin{equation}\label{iff_second_moment_1}
{\rm E}[\mathbf{\overline{y}}(k+1)\mathbf{\overline{y}}(k+1)^{\top}]
= {\rm E}[W(k) \mathbf{\overline{y}}(k)\mathbf{\overline{y}}(k)^{\top}W(k)].
\end{equation}}
{By applying the $\text{vec}(\cdot)$ operator to both sides, (\ref{iff_second_moment_1})  can be rewritten as}
\begin{equation}\label{iff_second_moment_2}
     {\rm E}[\mathbf{\overline{Y}}(k+1)]
    = {{\rm E}[W(k)\otimes W(k)]} {\rm E}[\mathbf{\overline{Y}}(k)],
\end{equation}
where $\mathbf{\overline{Y}}(k)=\text{vec}(\mathbf{\overline{y}}(k)\mathbf{\overline{y}}(k)^{\top}) \in \mathbb{R}^{n^2d^2}$. 
The convergence of the second moment, i.e., \[{\rm E}[\mathbf{\overline{y}}(k)\mathbf{\overline{y}}(k)^{\top}] \xrightarrow[]{k\rightarrow \infty} \mathbf{0}_{nd\times nd}\] is equivalent to ${\rm E}[\mathbf{\overline{Y}}(k) ]\xrightarrow[]{k\rightarrow \infty} \mathbf{0}_{n^2d^2}.$ {Note that vectorization is a linear transformation, it can be seen that
$\mathbf{1}_{n^2d^2}^\top\mathbf{\overline{Y}}(k) = \sum_{i,j=1}^n \text{vec}(\overline{y}_i^\top(k)\overline{y}_j(k))=\text{vec}\big(\sum_{i=1}^n \overline{y}_i(k) \sum_{j=1}^n \overline{y}_j(k)^\top\big)= 0$
, and (\ref{iff_second_moment_2}) thus has the same form as (\ref{lemma51}). Similar to the proof of \emph{Part 1}, all eigenvalues of ${\rm E}[W(k)\otimes W(k)-\frac{(\mathbf{1}_{n^2}\mathbf{1}_{n^2}^\top)\otimes\mathbf{I}_{d^2}}{n^2}]$ have to lie inside
the unit circle.}
This completes the proof of Lemma \ref{iff}.

\section{Proof of Theorem \ref{ddmwc}}
Denote the error vector $\overline{y}_i(k)=\overline{x}_i(k)-\overline{x}_0$. System (\ref{algorithm2}) can be rewritten as
\begin{equation}\label{B1}
    \begin{aligned}
    \overline{y}_i(k+1)&=\overline{y}_i(k)- \theta\alpha\sum_{j\in \mathcal{N}_i} \mathbf{A}_{ij}\big(\overline{y}_i(k)-\overline{y}_j(k)\big)\\
    &\qquad- (1-\theta)\alpha \mathbf{A}_{i0} \overline{y}_i(k).
    \end{aligned}
\end{equation}
For the whole system, (\ref{B1}) becomes
\begin{align} \label{B2}
    \mathbf{\overline{y}}(k+1)&=\big(\mathbf{I}_{dn}-\theta\alpha \mathbf{L}-(1-\theta)\alpha \text{blkdiag}({\mathbf{A}_{i0}})\big)\mathbf{\overline{y}}(k) \nonumber\\
    &=(\mathbf{I}_{dn}-\mathbf{N})\mathbf{\overline{y}}(k) =\mathbf{H}\mathbf{\overline{y}}(k),
\end{align}
where $\mathbf{\overline{y}}(k)=\text{vec}(\overline{y}_i(k))$.  

Choosing the updating step-sizes $\alpha$ satisfy Theorem \ref{ddmwc}, we have $\lambda(\alpha\mathbf{L})$, $\lambda(\alpha\text{blkdiag}({\mathbf{A}_{i0}})) \in [0,2)$ and thus the convex combination of these two matrices $\mathbf{N}$ satisfy $0\leq\lambda(\mathbf{N})<2$ and $-1<\lambda(\mathbf{H})\leq1$.
Under the assumptions (\ref{Ai0}) and (\ref{nullL}), $\mathbf{N}$ is positive definite (the complete proof can be found in \cite{Trinh2017ASCC}) and then $\lambda(\mathbf{H})<1$. As a result, it yields $-1<\lambda(\mathbf{H})<1$. 
The matrix $\mathbf{H}$ is thus stable, and $\mathbf{\overline{y}}(k)\rightarrow \mathbf{0}_{dn}$ as $k \rightarrow \infty$. Furthermore, it is well known from linear control theory that the converge rate of $\mathbf{\overline{y}}(k)$ will depend on the largest-in-magnitude eigenvalue of $\mathbf{H}$.

\section{Proof of Theorem \ref{2mmlf}}
Choosing the updating step sizes $\alpha$ satisfy Theorem \ref{2mmlf}, it is easily obtained that each possible $W_{ij}$ has eigenvalues satisfy $-1<\lambda(W_{ij})\leq 1$ and thus $0\leq\lambda(W_{ij}^\top W_{ij})\leq 1$. Denote $\left\{v_{ij}\right\}$ as the eigenspace of $W_{ij}$ corresponding to the eigenvalue $\lambda=1$. Clearly, $\left\{v_{ij}\right\}$ is also the eigenspace corresponding to the unity eigenvalue of $W_{ij}^\top W_{ij}$. We now treat the expectation ${\rm E}[W(k)]$ (resp., ${\rm E}[W(k)^\top W(k)]$)  as a convex combination of all possible $W_{ij}$ (resp., $W_{ij}^\top W_{ij}$) where ${\rm P}_{ij}\neq 0$. Because $W_{ij}$ is symmetric (and thus $W_{ij}^\top W_{ij}$), ${\rm E}[W(k)]$ cannot have a unity eigenvalue unless there exists a common eigenvector between every eigenspace $\left\{v_{ij}\right\}$. From Theorem \ref{ddmwc}, we already have $-1<\lambda({\rm E}[W(k)])<1$, which implies $\underset{{\rm P}_{ij}\neq 0}{\bigcap}\left\{v_{ij}\right\}=\varnothing$. Thus, it is obvious that $0\leq \lambda({\rm E}[W(k)^\top W(k)])<1$. This completes the proof of Theorem \ref{2mmlf}.

\section{Proof of Theorem \ref{NLmaintheorem}}
The proof is divided into two parts:

{\flushleft1) \emph{Convergence in Expectation.}} Taking the expectation of \eqref{algorithmNL13} yields ${\rm E}[W(k)]=\mathbf{I}_{{d}n_f}-\alpha\mathbf{L}^{\rm M}_{ff}$, 
which implies that
\begin{equation}\label{expcet}
    \begin{aligned}
        {\rm E}[\Tilde{\overline{\mathbf{x}}} _f(k+1)]
        &=(\mathbf{I}_{{d}n_f}-\alpha \mathbf{L}^{\rm M}_{ff}){\rm E}[\Tilde{\overline{\mathbf{x}}} _f(k)]\\
        &=(\mathbf{I}_{{d}n_f}-\alpha \mathbf{L}^{\rm M}_{ff})^{k+1}\Tilde{\overline{\mathbf{x}}} _f(0).
    \end{aligned}
\end{equation}
Similar to Appendix B, it could be easily obtained that by choosing the stepsize $\alpha <\underset{i \in V_f}{\min}\frac{1}{\max\| \mathbf{D}^{\rm M}_{i} \|}$, the following holds $-1<\lambda(\mathbf{I}_{{d}n_f}-\alpha \mathbf{L}^{\rm M}_{ff})\leq 1$. Noting that the matrix $\alpha \mathbf{L}^{\rm M}_{ff}$ has already been confirmed to be symmetric and positive definite. Hence, $-1<\lambda(\mathbf{I}_{{d}n_f}-\alpha \mathbf{L}^{\rm M}_{ff})<1$, i.e., system (\ref{expcet}) is exponentially stable. As a result, ${\rm E}[\Tilde{\overline{\mathbf{x}}} _f(k+1)]\rightarrow \mathbf{0}_{{dn_f}}$ as $k\rightarrow \infty$.

{\flushleft2) \emph{Convergence of the Second Moment.}} From the previous results, we obtain the following equation
\begin{equation}\label{secondmomentNL}
    \begin{aligned}
        {\rm E}[\Tilde{\overline{\mathbf{x}}}_f(k+1)^{\top}&\Tilde{\overline{\mathbf{x}}}_f(k+1)|\Tilde{\overline{\mathbf{x}}}_f(k)] \\
        &= \Tilde{\overline{\mathbf{x}}}_f(k)^{\top}{\rm E}[W(k)^{\top}W(k)]\Tilde{\overline{\mathbf{x}}}_f(k).
    \end{aligned}
\end{equation}

Hence, the proof of this part is similar to that of Theorem~\ref{2mmlf}.  Additionally, it can be easily obtained that the upper bound of $\epsilon$-consensus time of the proposed algorithm is as follows
$$K(\epsilon)=\frac{3{\log}(\epsilon^{-1})}{{\log}\lambda_{max}^{-1} \big( {\rm E}[{W}(k)^\top {W}(k)] \big)}.$$

\section{Proof of Theorem \ref{FC_convergence}}
By choosing $\alpha$ satisfying Theorem \ref{FC_convergence}, it is clearly seen that ${\rm E}[\mathbf{\overline{e}}_1(k)] \rightarrow \frac{1}{n}(\mathbf{1}_n\mathbf{1}_n^\top{\otimes \mathbf{I}_d })\mathbf{\overline{e}}_1(0)$, i.e., the average of initial position estimation errors (see Theorem \ref{upperbound1}). Moreover, since $\mathbf{I}_{dn}-\overline{K}$ is Schur followed by the choice of {$\sigma$}, the dynamic of the tracking error is stable. In order to find the expected tracking error in the steady state, one can take
\begin{equation}
 {\rm E}[\mathbf{\overline{e}}_2(\infty)]= (\mathbf{I}_{dn}-\overline{K}){\rm E}[\mathbf{{\overline{e}}}_2(\infty)]-\overline{K}{\rm E}[\mathbf{\overline{e}}_1(\infty)],   
\end{equation}
which is equivalent to
\begin{equation}
 {\rm E}[\mathbf{\overline{e}}_2(\infty)]= -{\rm E}[\mathbf{\overline{e}}_1(\infty)]=-\frac{1}{n}(\mathbf{1}_n\mathbf{1}_n^\top{\otimes \mathbf{I}_d })\mathbf{\overline{e}}_1(0).  
\end{equation}
Thus, the expectation of actual formation $\mathbf{\overline{x}}(k)$ will globally converges to $\mathbf{\overline{x}}^*-\frac{1}{n}(\mathbf{1}_n\mathbf{1}_n^\top{\otimes \mathbf{I}_d })\mathbf{\overline{e}}_1(0)$.

\bibliographystyle{IEEEtran}
\bibliography{ref}
\end{document}